%Paper: hep-ph/9305248
%From: HEHTH@SLACVM.SLAC.Stanford.EDU
%Date: 11 May 1993 18:18 -0800 (PST)

%macropackage=phyzzx

\def\unlock{\catcode`@=11} % This allows us to modify PLAIN macros.
\def\lock{\catcode`@=12} % at signs are no longer letters
\unlock
\def\refitem#1{\r@fitem{#1.}}
\lock
\refindent=20pt
\def\crr{\cr\noalign{\vskip 5pt}}
\def\us#1{\undertext{#1}}
\def\pri{^{\, \prime }}

\def\ls#1{\ifmath{_{\lower1.5pt\hbox{$\scriptstyle #1$}}}}
\def\9{\phantom 0}
\def\ifmath#1{\relax\ifmmode #1\else $#1$\fi}
\def\PRL#1&#2&#3&{\sl Phys.~Rev.~Lett.\ \bf #1\ \rm (19#2)\ #3}
\def\PRB#1&#2&#3&{\sl Phys.~Rev.\ \bf #1\ \rm (19#2)\ #3}
\def\NP#1&#2&#3&{\sl Nucl.~Phys.\ \bf #1\ \rm (19#2)\ #3}
\def\PRP#1&#2&#3&{\sl Phys.~Rep.\ \bf #1\ \rm (19#2)\ #3}
\def\PL#1&#2&#3&{\sl Phys.~Lett.\ \bf #1\ \rm (19#2)\ #3}
\def\wt{\widetilde}
\def\calm{{\cal M}}

\def\calv{{\cal V}}
\def\tb  {t_{\beta}}
\def\sw  {s_W}
\def\cw  {c_W}
\def\sb  {s_{\beta}}
\def\cb  {c_{\beta}}
\def\ctwob  {c_{2\beta}}
\def\sa  {s_{\alpha}}
\def\ca  {c_{\alpha}}
\def\sab  {s_{\alpha+\beta}}
\def\cab  {c_{\alpha+\beta}}
\def\sba  {s_{\beta-\alpha}}
\def\cba  {c_{\beta-\alpha}}
\def\tanb{\tan\beta}
\def\sinb{\sin\beta}
\def\cosb{\cos\beta}
\def\cosbma{\cos(\beta-\alpha)}
\def\sinbma{\sin(\beta-\alpha)}
\def\sina{\sin\alpha}
\def\cosa{\cos\alpha}
\def\sww{s_W^2}
\def\hl{h^0}
\def\hh{H^0}
\def\ha{A^0}
\def\mhl{m_{\hl}}
\def\mhh{m_{\hh}}
\def\mha{m_{\ha}}
\def\hpm{H^{\pm}}
\def\mhpm{m_{\hpm}}
\def\mhsm{m_{\phi^0}}
\def\hsm{\phi^0}
\def\mb{m_b}
\def\mt{m_t}
\def\mw{m_W}
\def\mzz{m_Z^2}
\def\mz{m_Z}
\def\mweak{M\ls{{\rm weak}}}
\def\msusy{M\ls{{\rm SUSY}}}
\def\msusyy{M\ls{{\rm SUSY}}^2}
\def\half{\ifmath{{\textstyle{1 \over 2}}}}
\def\threehalf{\ifmath{{\textstyle{3 \over 2}}}}
\def\fivehalf{\ifmath{{\textstyle{5 \over 2}}}}
\def\ninehalf{\ifmath{{\textstyle{9 \over 2}}}}
\def\twothirds{\ifmath{{\textstyle{2 \over 3}}}}
\def\sixteenthirds{\ifmath{{\textstyle{16\over 3}}}}
\def\twentythirds{\ifmath{{\textstyle{20 \over 3}}}}
\def\fortythirds{\ifmath{{\textstyle{40 \over 3}}}}
\def\sixtyfourthirds{\ifmath{{\textstyle{64 \over 3}}}}
\def\third{\ifmath{{\textstyle{1 \over 3}}}}
\def\fourth{\ifmath{{\textstyle{1\over 4}}}}
\def\ninefourth{\ifmath{{\textstyle{9 \over 4}}}}
\def\fifteenfourth{\ifmath{{\textstyle{15\over 4}}}}
\def\threefifths{\ifmath{{\textstyle{3 \over 5}}}}

\def\threeighth{\ifmath{{\textstyle{3 \over 8}}}}
\def\sevenninths{\ifmath{{\textstyle{7\over 9}}}}
\def\thirteenninths{\ifmath{{\textstyle{13\over 9}}}}
\def\fivetwelfth{\ifmath{{\textstyle{5 \over 12}}}}
\def\seventeentwelfth{\ifmath{{\textstyle{17 \over12}}}}
%%%%%%%%%%%%%%%%%%%%%%%%%%%%%%%%%%%%%%%%%%%%%%%%%%%%%%%%%%%%%
\Pubnum={SCIPP 93/06}
\date={March 1993}
\pubtype{}
\titlepage
\vsize=51pc
\singlespace
\parskip=2pt
\vskip4cm
\centerline{\fourteenbf When Are Radiative Corrections Important in}
\vskip5pt
\centerline{\fourteenbf the Minimal Supersymmetric Model?}
\vskip1cm
\centerline{\caps Howard E. Haber}
\centerline{\it Santa Cruz Institute for Particle Physics}
\centerline{\it University of California, Santa Cruz, CA 95064,
    U.S.A.}

\vskip1cm
\centerline{\caps Abstract}
\medskip
Precision electroweak measurements at LEP currently check the
validity of the Standard Model to about one part in a
thousand.  Any successful model of physics beyond the Standard Model
must be consistent with these observations.  The impact
of radiative corrections on the Minimal Supersymmetric Model (MSSM) is
considered.  The influence of supersymmetric
particles on precision electroweak measurements is generally
negligible since radiative corrections mediated by supersymmetric
particles are suppressed by a factor of order $\mz^2/\msusy^2$
(where $\msusy$ is the scale characterizing the scale of supersymmetric
particle masses).  However, there are a few pertinent exceptions.
For example,
the radiative corrections to the rare decay $b\to s\gamma$ from
charged Higgs and supersymmetric particle exchange can be of the
same order as the Standard Model contribution.
Large radiative corrections also lead to modifications
of MSSM tree-level (natural) relations.  The largest corrections
of this type occur in the MSSM Higgs sector and are enhanced
by powers of the top quark mass.  The consequences of the radiatively
corrected MSSM Higgs sector are briefly discussed.
\vfill
\centerline{Lecture given at the 23rd Wordshop of the INFN
Eloisatron Project}
\centerline{Erice, Italy, September 28--October 4, 1992}
\vfill
\endpage
\vskip1cm
\leftline{{\bf 1. Introduction}}
\medskip
\REF\susyrev{P. Fayet and S. Ferrara, {\sl Phys.~Rep.} {\bf 32} (1977)
249; H.P. Nilles, {\sl Phys.~Rep.} {\bf 110} (1984) 1;
H.E. Haber and G.L. Kane, {\sl Phys.~Rep.} {\bf 117} (1985) 75;
A.B. Lahanas and D.V. Nanopoulos, {\sl Phys. Rep.} {\bf 145} (1987) 1;
R. Barbieri, {\sl Riv. Nuovo Cimento} {\bf 11} (1988) 1.}
\REF\thooft{G. 't Hooft, in {\it Recent Developments in Gauge
Theories,} Proceedings of the NATO Advanced Summer Institute,
Cargese, 1979, edited by G. 't~Hooft \etal\ (Plenum, New York,
1980) p.~135.}
\REF\suss{L. Susskind, {\sl Phys. Rep.} {\bf 104} (1984) 181.}
\REF\susysol{E.~Witten, \sl Nucl.~Phys. \bf B188 \rm (1981) 513;
S.~Dimopoulos and H.~Georgi, \sl Nucl.~Phys. \bf B193 \rm (1981) 150;
N.~Sakai, \sl Z.~Phys. \bf C11 \rm (1981) 153; R.K. Kaul,
{\sl Phys. Lett.} {\bf 109B} (1982) 19.}
\REF\pdg{K. Hisaka \etal\ [Particle Data Group], \sl Phys. Rev. \bf
D45 \rm  (1992) S1.}
\REF\lepdata{The LEP Collaborations: ALEPH, DELPHI, L3 and OPAL,
{\sl Phys. Lett.} {\bf B276} (1992) 247.}
\REF\tatsu{M.E. Peskin and T. Takeuchi, {\sl Phys. Rev. Lett.} {\bf 65}
(1990) 964; {\sl Phys. Rev.} {\bf D46} (1992) 381.}
\REF\lepsearch{See, \eg, D.~Decamp \etal\ [ALEPH Collaboration],
{\sl Phys. Rep.} {\bf 216} (1992) 253.}
\REF\hhg{J.F. Gunion, H.E. Haber, G. Kane and S. Dawson,
{\it The Higgs Hunter's Guide} (Addison-Wesley Publishing Company,
Reading, MA, 1990) [Erratum: SCIPP-92/58 (1992)].}
Supersymmetric theories are considered by many theorists to be the
most likely theoretical framework to supersede the Standard Model of
elementary particles physics\refmark\susyrev.
The motivation for ``low-energy''
supersymmetry as a possible solution to the gauge hierarchy and
naturalness problems\refmark{\thooft--\susysol} is
well known and provides one reason that
many of us are participating in this meeting.  One
attractive feature of the simplest supersymmetric extensions of the
Standard Model is that it retains
elementary weakly coupled Higgs particles.  In particular, the Higgs
masses do not exceed the effective scale of supersymmetry breaking,
which is required by the consistency of the approach to be less than
of ${\cal O}(1~{\rm TeV})$.

At present, there is no direct experimental evidence for
supersymmetry\refmark\pdg.
Nevertheless, theorists who favor the supersymmetric approach
have been known (especially during an Ericean dinner) to
prepare lists of the
indirect evidence that give credence to the
supersymmetric universe.  In response,
a cynic might paraphrase Feynman's
famous dictum that a theory is suspect if it requires more than
one good reason
to convince the unconverted.  Nevertheless, it is fun to add to such a
hypothetical list of reasons to believe in supersymmetry.  Let me
contribute two reasons that will be elucidated in this
paper:
\pointbegin
Experimental deviations from the Standard Model in precision
electroweak measurements at LEP have not been detected\refmark\lepdata.

\noindent This is a clear confirmation\foot{Tongue slightly in
cheek.} of ``low-energy'' supersymmetry, where the typical
supersymmetric mass scale (denoted by $\msusy$) is above $\mz$.
The justification for this remark is based on the
considerations of section 3, where it is shown that supersymmetric
contributions to LEP electroweak observables are negligibly small
once $\msusy$ is somewhat larger than $\mz$.
In contrast, one finds the LEP measurements to be quite
constraining for many other models of physics
beyond the Standard Model\refmark\tatsu.
\point
The Higgs boson has not been discovered at LEP\refmark\lepsearch.

\noindent The Standard Model does not predict the mass
of the lightest Higgs boson\refmark\hhg.
In the minimal supersymmetric model,
radiative corrections raise the Higgs boson mass above its predicted
tree level range between 0 and $\mz$.  In section 6, we will see that
over a large range of
parameter space, the minimum value of the lightest Higgs mass lies
outside the current LEP Higgs mass limit, if the top quark mass is
sufficiently above $\mz$.

\REF\susygrand{U. Amaldi \etal, {\sl Phys. Rev.} {\bf D36} (1987) 1385;
U. Amaldi, W. de Boer and H. Furstenau, {\sl Phys. Lett.} {\bf B260}
(1991) 447; U. Amaldi \etal, {\sl Phys. Lett.} {\bf B281} (1992) 374.}
\REF\nano{J. Ellis, S. Kelly and D.V. Nanopoulos, {\sl Phys. Lett.}
{\bf B287} (1992) 95; {\sl Nucl. Phys.} {\bf B373} (1992) 55.}
\REF\zich{F. Anselmo, L. Cifarelli, A. Peterman, and A. Zichichi,
{\sl Nuovo Cim.} {\bf 104A} (1992) 1817; {\bf 105A} (1992) 581;
{\bf 105A} (1992) 1201.}
\REF\nonmin{J.R. Espinosa and M. Quiros, {\sl Phys. Lett.} {\bf B279}
(1992) 92; {\bf B302} (1993) 51;
G.L. Kane, C. Kolda, and J.D. Wells, UM-TH-92-24 (1992).}
In section 2, I review the parameters of the minimal supersymmetric
extension of the Standard Model (MSSM).  Although it is perhaps
dangerous to make sweeping generalizations based on such a minimal
model, two points should be kept in mind.  First, the unification
of the electroweak and strong coupling constants at around $10^{16}$~GeV
is a tantalizing clue
(and should be added to the above list!)
that low-energy supersymmetry may be correct.
In particular, using
renormalization group equations (RGEs) of the low-energy theory, one
finds that the coupling constants do not unify in the Standard Model,
while they do unify in a minimal supersymmetric model where $\msusy$ is
{\it roughly} 1 TeV (to within a couple of orders of magnitude)%
\refmark{\susygrand--\zich}.
Adding additional light Higgs doublets (beyond the minimal two doublets
required by the MSSM) destroys the unification.  Of course, other
non-minimal approaches cannot be ruled out, \eg, models with additional
Higgs singlets.  Second, in testing the supersymmetric approach, I
believe that the most sensible first step is to rigorously test
the most constrained model.  Here, the MSSM nicely fits the bill;
it is the simplest model with the fewest unknown parameters and hence
the most predictive.  It will be the first well-defined candidate
model for physics beyond the Standard Model to be either confirmed or
falsified by future experimental data.  Of course, it is  useful to
consider results of the supersymmetric approach that are independent
(or very weakly dependent) of the minimality assumption.  Some of
the results quoted in this paper are in fact more general than
the MSSM framework, while other results can be appropriately
generalized\refmark\nonmin.

Next, the influence of supersymmetric particle exchange
on various electroweak observables is described.
In section 3, I concentrate on the oblique radiative corrections---these
are corrections that arise due to the contributions of
new particles to gauge
boson self-energies.  I explain why all such corrections are
suppressed by a factor of order $\mz^2/\msusy^2$.  Numerically, I
demonstrate that the resulting oblique corrections mediated by
MSSM particles
are negligible once $\msusy$ is somewhat above $\mz$.  In section 4,
I briefly discuss non-oblique radiative corrections
induced by supersymmetric particle exchange.
As an example, deviations from the Standard Model
prediction for $b\to s
\gamma$ could be attributed to the existence of the charged Higgs boson
(which is required by the MSSM).  Finally, in sections 5 and 6, I focus
on the large radiative corrections to natural relations in the MSSM Higgs
sector.  Radiative corrections to MSSM Higgs masses and couplings
in the leading logarithmic approximation are described in section 5
and some numerical results are presented in section 6.  Finally,
section 7 concludes with a few summarizing remarks.
Some of the technical details are relegated to
three appendices.
\vskip1cm
\leftline{{\bf 2. The Minimal Supersymmetric Model (MSSM)}}
\medskip

%\REF\hehtasi{H.E. Haber, to appear in the \sl Proceedings of the
%1992 Theoretical Advanced Study Institute in Elementary Particle
%Physics \rm, Boulder, CO, 1--26 June 1992, SCIPP preprint (1992).}
\REF\hehpdg{H.E. Haber, {\sl Phys. Rev.} {\bf D45}, \rm
1 June 1992 Part II, p.~IX.5.}
\REF\ghi{J.F. Gunion and H.E. Haber, {\sl Nucl. Phys.} {\bf B272}
(1986) 1; {\bf B278} (1986) 449
[Erratum: UCD-92-31 and  SCIPP-92/59 (1992)].}
The Minimal Supersymmetric extension of the Standard Model (MSSM)
consists of taking the Standard Model as it is known today (including
the as yet undiscovered $t$-quark) and adding the corresponding
supersymmetric partners\refmark{\susyrev,\hehpdg}.
In addition, the MSSM must possess two Higgs doublets in
order to give masses to up and down type fermions in a manner consistent
with supersymmetry (and to avoid gauge anomalies introduced by
the fermionic superpartners of the Higgs bosons).  One then finds that
the MSSM Higgs sector is a CP-conserving two-Higgs doublet
model, whose Higgs potential is constrained by
supersymmetry\refmark\ghi.

The parameters of the MSSM fall into
two classes: a supersymmetry-conserving sector and
a supersymmetry-violating sector.
Among the parameters of the supersymmetry conserving
sector are: (i) gauge couplings: $g_s$, $g$ and $g^\prime$,
corresponding
to the Standard Model gauge group SU(3)$\times$SU(2)$\times$U(1)
respectively; (ii) Higgs Yukawa couplings: $\lambda_e$, $\lambda_u$, and
$\lambda_d$ (which are $3\times 3$ matrices in flavor space); and
(iii) a supersymmetry-conserving Higgs mass parameter $\mu$.  The
supersymmetry-violating sector contains the following set of parameters:
(i) gaugino Majorana masses $M_3$, $M_2$ and $M_1$ associated with
the SU(3), SU(2) and U(1) subgroups of the Standard Model;
(ii) scalar mass matrices for the squarks and sleptons; (iii)
Higgs-squark-squark trilinear interaction terms (the so-called
``A-parameters'') and corresponding terms involving the sleptons;
and (iv) three scalar Higgs mass parameters---two
diagonal and one off-diagonal mass terms for the two Higgs doublets.
These three mass parameters can be re-expressed in terms of the two
Higgs vacuum expectation values, $v_1$ and $v_2$, and one physical Higgs
mass.  Here, $v_1$ ($v_2$) is the vacuum
expectation value of the Higgs field which couples exclusively
to down-type (up-type) quarks and leptons.  Note that $v_1^2+v_2^2=
(246~{\rm GeV})^2$ is fixed by the $W$ mass while the ratio
$\tan \beta = v_2/v_1$ is a free parameter of the model.

\REF\inos{Explicit forms for the chargino and neutralino mass matrices
can be found in Appendix A of ref.~[\ghi].}
The supersymmetric partners of the electroweak gauge
and Higgs bosons (the gauginos and higgsinos)
can mix.  As a result,
the physical mass eigenstates are model-dependent linear combinations
of these states, called charginos and neutralinos, which
are obtained by diagonalizing the corresponding mass
matrices\refmark\inos.
The chargino mass matrix depends on $M_2$, $\mu$, $\tan\beta$ and
$m_W$, while the
neutralino mass matrix depends on $M_1$, $M_2$, $\mu$,
$\tan\beta$, $m_Z$ and the weak mixing angle $\theta_W$.
It is common practice in the literature to reduce the
supersymmetric parameter freedom
by requiring that all three gaugino mass parameters
are equal at some grand unification scale.  Then,
at the electroweak scale, the gaugino mass parameters can be expressed in
terms of the gluino mass, $|M_3|$. The other
two gaugino mass parameters are given by
$M_2=(g^2/g_s^2)M_3$ and
$M_1=(5g^{\prime\,2}/3g^2)M_2$.
Having made this assumption, the chargino and neutralino masses and
mixing angles depend only on three unknown parameters: the gluino mass,
$\mu$, and $\tan\beta$.

\REF\nath{P.~Nath, R.~Arnowitt and A.~H.~Chamseddine, {\it Applied
$N=1$ Supergravity} (World Scientific, Singapore, 1984).}
The supersymmetric partners of the quarks and leptons are the spin-zero
squarks and sleptons.  For a given fermion
$f$, there are two supersymmetric partners $\widetilde
f_L$ and $\widetilde f_R$ which are scalar partners of the
corresponding left and
right-handed fermion.  (There is no $\widetilde\nu_R$.)
For simplicity, I shall ignore intergenerational mixing.  Then the
squark and slepton masses of one generation are obtained from
mass matrices with both diagonal and off-diagonal elements (the latter
correspond to
$\widetilde f_L$-$\widetilde f_R$ mixing).  Using first generation
notation, the (diagonal) $L$ and $R$-type
squark and slepton masses are given by\refmark\nath\
$$\eqalign{%
  M^2_{\tilde u_L} &= M^2_{\wt Q}+m_u^2+m_Z^2\cos 2\beta(\half-
\twothirds \sin^2\theta_W) \cr
  M^2_{\tilde u_R} &=M^2_{\wt U}+m_u^2+\twothirds m_Z^2\cos 2\beta
\sin^2\theta_W \cr
  M^2_{\tilde d_L} &=M^2_{\wt Q}+m_d^2-m_Z^2\cos 2\beta(\half-\third
\sin^2\theta_W) \cr
   M^2_{\tilde d_R} &=M^2_{\wt D}+m_d^2-\third m_Z^2\cos 2\beta
\sin^2\theta_W \cr
   M^2_{\tilde \nu}\; &=M^2_{\wt L}+\half m_Z^2\cos 2\beta\cr
   M^2_{\tilde e_L} &=M^2_{\wt L}+m_e^2-m_Z^2\cos 2\beta(\half-
\sin^2\theta_W) \cr
   M^2_{\tilde e_R} &=M^2_{\wt E}+m_e^2-m_Z^2\cos 2\beta\sin^2\theta_W\,.
\cr }\eqn\diagsqmasses$$
\REF\rudaz{J.~Ellis and S.~Rudaz, \sl Phys.~Lett. \bf 128B, \rm 248
(1983).}
The soft-supersymmetry-breaking parameters $M_{\wt Q}$,
$M_{\wt U}$, $M_{\wt D}$, $M_{\wt L}$ and $M_{\wt E}$ are
unknown parameters.  In this paper, we will consider the simplest
case, where all such unknown mass parameters are taken to be equal
to one common mass, $\msusy$.  The strength of the
$\widetilde f_L$---$\widetilde f_R$ mixing is proportional
to the corresponding off-diagonal element of the scalar
mass-squared-matrix\refmark\rudaz\
$${M_{LR}^2=\cases{m_d(A_d-\mu\tan\beta),&for ``down''-type $f$\cr
       m_u(A_u-\mu\cot\beta),&for ``up''-type $f$,\cr}}\eqn\mlr$$
where $m_d$ ($m_u$) is the mass of the appropriate
``down'' (``up'') type quark or lepton.
Here, $A_d$ and $A_u$ are
soft-supersymmetry-breaking $A$--parameters and $\mu$ and $\tan\beta$
have been defined earlier.  Due to the appearance
of the {\it fermion} mass in eq.~\mlr, one expects $M_{LR}$ to be small
compared to the diagonal squark and slepton masses, with the possible
exception of the top-squark, since $m_t$ is large.

\REF\hhgref{For a comprehensive review and a complete set of references,
see Chapter 4 of ref.~[\hhg].}
Finally, consider the Higgs sector of the MSSM\refmark\hhgref.
Although this is not a supersymmetric sector, supersymmetry imposes
strong constraints on the form of the quartic
couplings of the Higgs fields.  Using the notation of Appendix
A, the quartic couplings $\lambda_i$ are given by
$$\eqalign{%
\lambda_1 &=\lambda_2 = \fourth (g^2+g'^2)\,,\cr
\lambda_3 &=\fourth (g^2-g'^2)\,,\cr
\lambda_4 &=-\half g^2\,,\cr
\lambda_5 &=\lambda_6=\lambda_7=0\,.\cr}
\eqn\bndfr$$
Inserting these results into eqs.~\mamthree\ and \massmhh, it
follows that
$$
\mha^2 =m_{12}^2(\tan\beta+\cot\beta)\,,
\eqn\susymha$$
$$
\mhpm^2 =\mha^2+\mw^2\,,\phantom{\cot\beta))}
\eqn\susymhpm$$
and the neutral CP-even mass matrix is given by
$$
{\cal M}^2 =
   \pmatrix{\mha^2 \sin^2\beta + m^2_Z \cos^2\beta&
 -(\mha^2+m^2_Z)\sin\beta\cos\beta\cr -(\mha^2+m^2_Z)\sin\beta\cos\beta&
         \mha^2\cos^2\beta+ m^2_Z \sin^2\beta \cr }\,.\eqn\kv
$$
The eigenvalues of ${\cal M}^2$ are
the squared masses of the two CP-even Higgs scalars
$$
  m^2_{H^0,h^0} = \half \left( \mha^2 + m^2_Z \pm
                  \sqrt{(\mha^2+m^2_Z)^2 - 4m^2_Z \mha^2 \cos^2 2\beta}
                  \; \right)\,.\eqn\kviii
$$
and the diagonalizing angle is $\alpha$, with
$$
  \cos 2\alpha = -\cos 2\beta \left( {\mha^2-m^2_Z \over
                  m^2_{H^0}-m^2_{h^0}}\right)\,,\qquad
  \sin 2\alpha = -\sin 2\beta \left( m^2_{H^0} + m^2_{h^0} \over
                   m^2_{H^0}-m^2_{h^0} \right)\,.\eqn\kix
$$
{}From the expressions for the Higgs masses obtained above,
the following inequalities are easily established
$$\eqalign{
  m_{h^0} &\leq \mha \cr
  m_{h^0} &\leq m|\cos 2\beta | \leq m_Z \,,
            \qquad {\rm with}\ m \equiv min(m_Z,\mha) \cr
  m_{H^0} &\geq m_Z\,,\cr   m_{H^\pm} &\geq\mw\,.\cr }\eqn\kx
$$
Eqs.~\susymhpm-\kx are tree-level results which are modified
when radiative corrections are incorporated.  These will be discussed
in detail in section 5.

\FIG\tanlima{%
The region of $\tanb$--$\mt$ parameter space in which
all running Higgs-fermion Yukawa couplings remain finite at all
energy scales, $\mu$, from $\mz$ to
$\Lambda=10^{16}$~GeV\refmark\handz.
Non-supersymmetric two-Higgs-doublet (one-loop)
renormalization group equations \break
(RGEs) are used for $\mz\leq\mu\leq\msusy$ and the RGEs of the
minimal supersymmetric model are used for $\msusy\leq\mu\leq\Lambda$
(see Appendix B).  Five different values of $\msusy$ are shown;
the allowed parameter space lies below the respective curves.}
\FIG\tanlimb{%
The region of $\tanb$--$\mt$ parameter space in which
all running Higgs-fermion Yukawa couplings remain finite at all
energy scales from $\mz$ to
$\Lambda=100$ TeV.  See caption to fig.~\tanlima.}
%\pageinsert
%   \tenpoint \baselineskip=12pt   \narrower
%\plotpicture{\hsize}{8cm}{tanbl.topdraw}
%          \noindent
%{\bf Fig.~\tanlima.}\enskip
%
%\vfill
%\plotpicture{\hsize}{8cm}{tanbl2.topdraw}
%          \noindent
%{\bf Fig.~\tanlimb.}\enskip
%\endinsert

\REF\alef{D. Decamp \etal\ [ALEPH Collaboration], {\sl Phys. Lett.}
{\bf B265} (1991) 475; D. Buskulic \etal\ [ALEPH Collaboration],
{\sl Phys. Lett.} {\bf B285} (1992) 309.}%
\REF\physlet{R. Hempfling, {\sl Phys. Lett.} {\bf B296} (1992) 121.}
%\REF\hewett{V. Barger, J.L. Hewett and R.J.N. Phillips, {\sl Phys.
%Rev.} {\bf D41} (1990) 3421.}
\REF\tanbt{See \eg, G.F. Giudice and G. Ridolfi, {\sl Z. Phys.}
{\bf C41} (1988) 447; M. Olechowski and S. Pokorski, {\sl
Phys. Lett.} {\bf B214} (1988) 393; M. Drees and M.M. Nojiri,
{\sl Nucl. Phys.} {\bf B369} (1992) 54.}
\REF\bagger{J. Bagger, S. Dimopoulos and  E. Masso, {\sl Phys. Lett.}
{\bf 156B} (1985) 357; \sl Phys. Rev. Lett. \bf 55\rm (1985) 920.}
\REF\russ{G.M. Asatryan, A.N. Ioannisyan and S.G. Matinyan,
{\sl Sov. J. Nucl. Phys.} {\bf 53} (1991) 371 [{\sl Yad. Fiz.} {\bf 53},
(1991) 592];
M. Carena, T.E. Clark, C.E.M. Wagner, W.A. Bardeen and K. Sasaki,
{\sl Nucl. Phys.} {\bf D369} (1992) 33.}
\REF\handz{H.E. Haber and F. Zwirner, unpublished.}
\REF\cdflimit{F. Abe \etal\ [CDF Collaboration], {\sl Phys. Rev. Lett.}
{\bf 68} (1992) 447.}\
It is clear from the above results that the MSSM Higgs sector
tree-level masses and couplings are determined in terms of two
parameters.  It is convenient to choose these two parameters to be
the mass of the CP-odd Higgs boson, $\mha$, and
the ratio of vacuum expectation values, $\tanb$.  The mass $\mha$
depends on the (unknown) parameter $m_{12}^2$ [see
eq.~\susymha] which is a soft-supersymmetry-breaking mass.
It is tempting to take this parameter of order $\msusy$.  In this
case, for $\msusy\gg\mz$, the $\hpm$, $\hh$ and $\ha$ would all
have large masses, and the Higgs sector of the effective low-energy
theory below $\msusy$ would correspond precisely to the minimal Higgs
sector of the Standard Model.  In this paper, I shall be a little
more flexible and permit any value of $\mha$ from 0 to
${\cal O}(\msusy)$.\foot{Actually, present LEP limits from the ALEPH
collaboration\refmark\alef\
suggest that $\mha>20$ GeV (at 95\% CL). However,
the limit on $\mhl$ may be substantially weaker if large
squark mixing is permitted\refmark\physlet.}
Limits on the parameter $\tanb$ can be obtained based on
theoretical considerations.  For example,
models of low-energy supersymmetry based on supergravity
strongly favor $\tanb>1$\refmark\tanbt.  Even in the absence of
specific models, values of $\tanb<1$ are theoretically disfavored
because once $\tanb$
becomes too small, the Higgs couplings to top quarks become strong.
%[see eqs.~\qqcouplings--\hpmqq].
%Eventually, the tree-unitarity of processes involving the
%Higgs-top quark Yukawa
%coupling is violated.  Perhaps this should not be
%regarded as a theoretical defect, although it does render any
%perturbative analysis unreliable.  A rough lower bound advocated
%by ref.~\hewett, $\tanb\gsim m_t/600$~GeV, corresponds to a Higgs-top
%quark coupling in the perturbative region.  A similar argument
%involving the Higgs-bottom quark coupling would yield $\tanb\lsim 120$.
%A more solid theoretical constraint is based on the
By imposing the
requirement that the Higgs--fermion couplings remain finite
when running from the electroweak scale to
some large energy scale $\Lambda$\refmark{\bagger-\handz},
one obtains limits on $\tanb$ as a function of $\mt$ and $\Lambda$.%
\foot{Beyond the energy scale
$\Lambda$, one assumes that new physics enters.}
Using the renormalization group equations given in Appendix B,
we integrate from the electroweak scale to $\Lambda$ (allowing for
the possible existence of a supersymmetry-breaking
scale in the range $\mz\leq\msusy\leq
\Lambda$), and determine the region of $\tanb$--$\mt$ parameter space
in which the Higgs-fermion Yukawa couplings remain finite.
(The $t$, $b$ and $\tau$ are all included in
the analysis.)  The results are shown in figs.~\tanlima\ and
\tanlimb\ for two different choices of $\Lambda$\refmark\handz.
The allowed region of parameter space lies below the curves shown.
For example, if there is no new physics (other than perhaps minimal
supersymmetry) below the grand unification scale of $10^{16}$~GeV,
then based on the CDF limit\refmark\cdflimit\ of $\mt>91$~GeV, one
would conclude that $0.5\lsim\tanb\lsim 50$.  The lower limit on
$\tanb$ becomes even sharper if the top-quark mass is heavier.
Remarkably, the limits on $\tanb$ do not get substantially weaker
for $\Lambda$ as low as 100 TeV.

\bigskip
\leftline{{\bf 3. MSSM Contributions to Precision Electroweak
Measurements}}
\medskip

\REF\barbi{R. Barbieri, M. Frigeni, F. Giuliani and H.E. Haber,
{\sl Nucl. Phys.} {\bf B341} (1990) 309; M. Drees and K. Hagiwara,
{\sl Phys. Rev.} {\bf D42} (1990) 1709;
P. Gosdzinsky and J. Sola, {\sl Phys. Lett.} {\bf B254} (1991) 139;
{\sl Mod. Phys. Lett.} {\bf A6} (1991) 1943; M. Drees, K. Hagiwara,
and A. Yamada, {\sl Phys. Rev.} {\bf D45} (1992) 1725;
R. Barbieri, M. Frigeni and F. Caravaglios, {\sl Phys. Lett.}
{\bf B279} (1992) 169; J. Ellis, G.L. Fogli, and E. Lisi,
{\sl Phys. Lett.} {\bf B286} (1992) 85; CERN-TH-6643 (1992).}
\REF\hehth{H.E. Haber, SCIPP preprint in preparation.}
The Standard Model of particle physics
provides a detailed and accurate description of all
observed high energy physics phenomena to date.  Moreover,
precision measurements of electroweak observables at LEP are now
becoming sensitive to the one-loop predictions of the theory.
Apart from providing very sensitive tests of the Standard Model,
such measurements can impose severe constraints on theories that
incorporate physics beyond the Standard Model.  In this section we
examine the impact of the MSSM on the one-loop electroweak radiative
corrections\refmark{\barbi,\hehth}.

\REF\decouple{T. Appelquist and J. Carazzone, {\sl Phys. Rev.} {\bf
D11} (1975) 2856; B. Ovrut and H. Schnitzer, {\sl Phys. Rev.} {\bf
D22} (1980) 2518; Y. Kazama and Y.-P. Yao, {\sl Phys. Rev.} {\bf D25}
(1982) 1605; J.C. Collins, {\it Renormalization} (Cambridge University
Press, Cambridge, 1984), Chapter 8.}
\REF\veltman{M. Veltman, {\sl Nucl. Phys.} {\bf B123} (1977) 89;
M.B. Einhorn, D.R.T. Jones, and M. Veltman, {\sl Nucl. Phys.}
{\bf B191} (1981) 146.}
\REF\nondecouple{J.C. Collins, F. Wilczek and A. Zee, {\sl Phys. Rev.}
{\bf D18} (1978) 242.}
\REF\doug{%
D. Toussaint, {\sl Phys. Rev.} {\bf D18} (1978) 1626.}
\REF\cfh{M. Chanowitz, M.
Furman and I. Hinchliffe, {\sl Phys. Lett.} {\bf 78B}
(1978) 285; {\sl Nucl. Phys.} {\bf B153} (1979) 402.}
\REF\sirlin{%
W.J. Marciano and A. Sirlin, {\sl Phys. Rev.} {\bf D22} (1980) 2695
[E: {\bf D31} (1985) 213].}
\REF\abj{G. Altarelli, R. Barbieri, and S. Jadach, {\sl Nucl. Phys.}
{\bf B369} (1992) 3 [E: {\bf B376} (1992) 444].}
\REF\lang{P. Langacker and M. Luo, {\sl Phys. Rev.} {\bf D44} (1991)
817; J. Ellis, G.L. Fogli
and E. Lisi, {\sl Phys. Lett.} {\bf B274} (1992) 456;
D. Schaile, {\sl Z. Phys.}
{\bf C54} (1992) 387; P. Renton, {\sl Z. Phys.} {\bf C56} (1992) 355.}
\REF\paul{P. Langacker, in {\it Electroweak Physics Beyond the
Standard Model}, Proceedings of the International Workshop on Electroweak
Interaction Beyond the Standard Model, Valencia, Spain, Oct 2-5, 1991,
edited by  J.W.F. Valle and J. Velasco (World Scientific, Singapore,
1992) p.~75.}
\REF\screen{M. Veltman, {\sl Acta Phys. Pol.} {\bf B8} (1977) 475;
M.B. Einhorn and J. Wudka, {\sl Phys. Rev.} {\bf D39} (1989) 2758.}

The Standard Model is usually
assumed to be an effective low-energy limit
of a more fundamental theory.  Let $M$ be the minimum
characteristic energy scale at which new physics beyond the Standard
Model enters.  For example, theoretical attempts to address the
hierarchy and naturalness problems usually take
$M$ no larger than about 1 TeV.  We then can pose the following
question.  Can evidence of physics at the scale $M$ be detected
through its virtual corrections to low-energy electroweak observables?
In order to address this question, one must first determine the
expected size of such radiative corrections.  The
decoupling theorem\refmark\decouple\
implies that radiative corrections to
electroweak observables from such new physics
should be of ${\cal O}(g^2\mz^2/M^2)$.
That is, the virtual effects due to new physics formally decouple as
$M\to\infty$.  However, in spontaneously broken gauge theories, an
exception to the decoupling theorem
arises\refmark{\veltman--\doug}.
Suppose we wish to
consider the virtual effects of a certain particle whose mass $m$
is proportional to a dimensionless coupling of the theory.
In this case,
the virtual effects due to such a particle do {\it not} decouple in the
large $m$ limit.
Two examples in the Standard Model are the top quark, whose mass
is proportional to a Higgs-quark Yukawa coupling, and the Higgs boson,
whose mass is proportional to the Higgs self-coupling.

The classic example of non-decoupling can be seen in the $\rho$-parameter
of electroweak physics\refmark{\veltman,\cfh}.
If one defines $\rho\equiv\mw^2/\mz^2\cos^2\theta_W$, then at tree-level
$\rho=1$ in any SU(2)$\times$U(1) model whose
Higgs sector consists entirely of weak scalar doublets (\eg, the
Standard Model and the MSSM).  Since $\rho=1$ is a ``natural'' relation
in such models, the deviation of $\rho$ from one is calculable when
radiative corrections are incorporated.  To proceed, one must
carefully define the observables of the model in order to establish
a useful one-loop definition for $\rho$.  For example, in ref.~[\sirlin],
$\rho\equiv\rho\ls{NC}$
is defined as the renormalization factor in the one-loop neutral current
neutrino-nucleon scattering amplitudes when expressed in terms of the
Fermi constant $G\ls{F}$.
A recent theoretical analysis of LEP data in ref.~[\abj] gives
$\rho= 0.9995\pm 0.0051$.
In the Standard Model, the deviation of
$\rho$ from 1 is predicted to be
quite small unless there exist particles with
nontrivial electroweak quantum numbers that are substantially
heavier than the $Z$ boson.
It is therefore convenient to define
$$\rho\equiv \rho\ls{\rm RSM}+\delta\rho\,,\eqn\refsm$$
where $\rho\ls{\rm RSM}\simeq 1$ is the $\rho$ parameter in a
``reference Standard Model'' (RSM) in which the radiative corrections
to $\rho$ are very small.
In order to exhibit the explicit
dependence of $\rho$ on the top quark and Higgs masses, let us
choose a RSM in which $\mt=\mb$ and $\mhsm=\mz$
(where $\mhsm$ is the mass of the Standard Model Higgs boson).
Then, in the Standard Model,
assuming $\mhsm\gg\mz$\refmark{\veltman,\cfh,\sirlin}
$$
  \delta\rho \simeq{g^2N_c\over 32\pi^2 m^2_W}F(\mt^2,\mb^2)
          -{3 g^2 \over 64 \pi^{2} c\ls{W}^2}
\left[s^2\ls W\ln \biggl(
 {\mhsm^2 \over \mz^2} \biggr)-{c\ls{W}^4\over s\ls{W}^2}\ln\left(
{\mw^2\over\mz^2}\right)-1\right]\,,
\eqn\rhocorrection$$
where $N_c=3$,
$s\ls{W}\equiv\sin\theta_W$, $c\ls{W}\equiv\cos\theta_W$, and the
function $F$ is defined in eq.~\effdef.
In deriving eq.~\rhocorrection,
$\mhsm$ is assumed to be much larger than $\mz$.
As advertised, the decoupling theorem is not respected in the
limit of large top quark and/or Higgs mass.  The quadratic dependence
of $\delta\rho$ on $m_t$ yields a useful constraint on the top quark
mass.\foot{Radiative corrections to other electroweak observables
also depend nontrivially on $m_t$ (\eg, see refs.~[\abj--\paul]).
A comprehensive analysis of electroweak data quoted in
the 1992 Particle Data Group compilation\refmark\pdg\
yields $\mt<201$~GeV at $95\%$~CL.}
In contrast, the dependence of $\delta\rho$
on Higgs mass is only logarithmic and is therefore not very useful in
constraining the Higgs mass.\foot{This is an example of Veltman's
screening theorem\refmark\screen.}
\REF\oblique{B.W. Lynn, M.E. Peskin and R.G. Stuart, in {\it Physics at
LEP}, edited by J. Ellis and R. Peccei, CERN Yellow Report CERN-86-02
(1986) p.~90;
B. Lynn and D.C. Kennedy, {\sl Nucl. Phys.} {\bf B321}
(1989) 83; {\bf B322} (1989) 1; D.C. Kennedy, {\sl Nucl. Phys.} {\bf
B351} (1991) 81; M.E. Peskin, in {\it Physics at the 100~GeV Mass
Scale}, Proceedings of the 1989 SLAC Summer Institute, edited by
E.C. Brennan, SLAC-Report-361 (1990) p.~71.}

The contributions of physics beyond the Standard Model to electroweak
observables occur primarily through virtual loop corrections to
gauge boson propagators, sometimes called ``oblique''
corrections\refmark\oblique.  As an example, this is typically a
good approximation in the case
of virtual Higgs boson corrections, since the Higgs coupling to
light fermions is suppressed by a factor $m_f/\mw$.  However,
there are a number of cases where one-loop vertex and box
corrections involving the
coupling of charged Higgs bosons to $t\bar b$ are not especially small.
I will consider this possibility briefly in section 4.
In this section, I shall work in the oblique approximation and assume
that the radiative corrections to an electroweak observable of
interest are dominated by
virtual heavy particle corrections to gauge boson
propagators.

\REF\otherst{G. Altarelli and R. Barbieri, {\sl Phys. Lett.} {\bf B253}
(1990) 161.}
\REF\morest{W.J. Marciano and J.L. Rosner, {\sl Phys. Rev. Lett.}
{\bf 65} (1990) 2963;
D.C. Kennedy and P. Langacker, {\sl Phys. Rev. Lett.} {\bf 65} (1990)
2967 [E: {\bf 66} (1991) 395]; {\sl Phys. Rev.} {\bf D44} (1991) 1591;
D.C. Kennedy, {\sl Phys. Lett.}, {\bf B268} (1991) 86; J.Ellis,
G.L. Fogli, and E. Lisi, {\sl Phys. Lett.} {\bf B285} (1992) 238.}
\REF\kennedy{D.C. Kennedy, in {\it Perspectives in the Standard Model},
Proceedings of the 1991 Theoretical Advanced Study Institute in
Elementary Particle Physics, Boulder, CO, 3-28 June 1991,
edited by R.K. Ellis, C.T. Hill, and J.D. Lykken (World Scientific,
Singapore, 1992) p.~163.}
In the limit where the heavy
particle masses are much larger than the $Z$ mass, the heavy particle
contributions to oblique radiative corrections can be summarized in
terms of three numbers called $S$, $T$ and $U$\refmark{\tatsu,\otherst--%
\kennedy}.
$T$ is related simply to the $\rho$ parameter
$$\rho-1=\alpha T\,,\eqn\tdef$$
where $\alpha$ is the usual fine structure constant.  To formally
define the three quantities $S$, $T$ and $U$, one proceeds as follows.
Let
$$i\Pi^{\mu\nu}_{ij}(q)=ig^{\mu\nu}A_{ij}(q^2)+iq^\mu q^\nu
B_{ij}(q^2)\,,\eqn\propagator$$
be the sum of all one-loop Feynman graphs contributing to
the $V_i$--$V_j$ two-point function, where $q$ is the
four-momentum of the vector boson ($V=W,\ Z$ or $\gamma$).
Only the functions $A_{ij}$ are
relevant for the subsequent analysis.  It is convenient to write
$$A_{ij}(q^2)=A_{ij}(0)+q^2F_{ij}(q^2)\,,\eqn\fdef$$
which define the quantities $F_{ij}$.  Gauge invariance
implies that $A_{\gamma\gamma}(0)=0$.%
\foot{In addition, the sum of
heavy particle contributions to $A_{Z\gamma}(0)$ also vanishes exactly.
Only gauge boson loops can produce
nonzero contributions to $A_{Z\gamma}(0)$ (in the standard $R$-gauge).}
A major simplification takes place if one is interested in the
effects of ``heavy'' physics (characterized by a scale $M\gg\mz$)
on electroweak observables.  In this case, since
$q^2$ is of order $\mz^2$,  one only makes
an error of ${\cal O}(\mz^2/M^2)$ by neglecting the $q^2$ dependence
of the $F_{ij}$.  Then, one can show that the oblique corrections
to electroweak observables due to heavy physics can be expressed
in terms of three particular
combinations of the $A_{ij}(0)$ and $F_{ij}$
$$\eqalign{\alpha T&\equiv {A\ls{WW}(0)\over\mw^2}-{A\ls{ZZ}(0)\over
\mz^2}-{2s\ls{W}\over c\ls W}{A_{Z\gamma}(0)\over\mz^2}\cr
{g^2\over 16\pi c\ls{W}^2}S&\equiv F\ls{ZZ}(\mz^2)
-F\ls{\gamma\gamma}(\mz^2)+
\left({2s\ls{W}^2-1\over s\ls{W} c\ls{W}}\right)F\ls{Z\gamma}(\mz^2)\cr
{g^2\over 16\pi}(S+U)&\equiv  F\ls{WW}(\mw^2)-F\ls{\gamma\gamma}(\mw^2)-
{c\ls{W}\over s\ls{W}}F\ls{Z\gamma}(\mw^2)\,.\cr}
\eqn\studefs$$
Note that the $A_{ij}(0)$ and $F_{ij}$ in the above formulae
are divergent quantities.  Nevertheless, if one includes a complete
set of contributions from a gauge invariant sector, then $S$, $T$,
and $U$ will be finite constants.
The Higgs sector by itself does not constitute a gauge invariant
sector in this regard, so one must include the vector boson sector
as well to obtain a non-divergent result for $S$, $T$ and $U$.
Alternatively, in order to obtain finite quantities that solely
reflect the influence of heavy Higgs physics, one can
define $\delta S$, $\delta T$ and $\delta U$ relative to some
reference Standard Model  where $\mhsm$ is fixed to a convenient
value.  For example, if we choose a RSM with
$\mhsm=\mz$, then the
change in $S$ due to a fourth generation of fermions $U$ and $D$ (with
electric charges $e\ls{D}+1$ and $e\ls{D}$ respectively) and a heavy
Higgs boson of mass $\mhsm$ is given by
$$
\delta S\simeq {N_c\over 6\pi}\left[1+(1+2e\ls{D})\ln\left(
{m\ls{D}^2\over m\ls{U}^2}\right)\right]
+{1\over 12\pi}\left[\ln\left({\mhsm^2\over\mz^2}\right)-3\pi\sqrt{3}
+{107\over 6}\right]\,,\eqn\esscorrection$$
where $m\ls{U}$, $m\ls{D}$, $\mhsm\gg\mz$ has been assumed.
Once again, the non-decoupling effects of the heavy physics are
apparent.

Radiative corrections (in the oblique approximation) to electroweak
observables can be expressed in terms of $S$, $T$ and $U$.
The $\rho$-parameter discussed above
is one such example.  A second example is the $W$ mass prediction.
The one-loop prediction is obtained by solving the following equation
for the $W$ mass
$$\mw^2\left(1-{\mw^2\over\mz^2}\right)=\left({\pi\alpha\over\sqrt{2}
G\ls{F}}\right)^2{1\over 1-\Delta r}\,,\eqn\wmass$$
\vskip6pt\noindent
where $\pi\alpha\big/\sqrt{2}G\ls{F}=(37.2802~{\rm GeV})^2$ and
\vskip6pt
$$\Delta r={g^2\over 8\pi}\left[S-2c\ls{W}^2 T+\left({2s\ls{W}^2-1\over
2s\ls{W}^2}\right)U\right]\,.\eqn\deltar$$
\vskip6pt\noindent
Other examples can be found in refs.~[\tatsu,\otherst--\morest].
Thus the effects of heavy physics on numerous
electroweak observables are immediately known once
the corresponding
contributions to $S$, $T$ and $U$ have been computed.

\REF\higgsrad{S. Bertolini, {\sl Nucl. Phys.} {\bf B272} (1986) 77.}
\REF\oldhiggsrad{R.S. Lytel, {\sl Phys. Rev.} {\bf D22} (1980) 505;
J.-M. Fr\`ere and J.A.M. Vermaseren, {\sl Z. Phys.} {\bf C19} (1983) 63;
W. Hollik, {\sl Z. Phys.} {\bf C32} (1986) 291; {\bf C37} (1988) 569.}
\REF\higgsradtwo{C.D. Froggatt, R.G. Moorhouse, and I.G. Knowles,
{\sl Phys. Rev.} {\bf D45} (1992) 2471; {\sl Nucl. Phys.} {\bf B386}
(1992) 63; T. Inami, C.S. Lim and A. Yamada, {\sl Mod. Phys. Lett.}
{\bf A7} (1992) 2789.}

In computing the contributions
of the MSSM to $S,\ T$ and $U$, it is convenient to consider
separately the contributions from the
various sectors of particles not contained in the Standard Model to
the gauge bosons self-energies.  Each sector, when appropriately
defined, yields a finite shift to $S$, $T$ and $U$.
The relevant particle sectors include all supersymmetric
particles and the physical Higgs bosons (beyond the minimal
neutral Higgs scalar of the Standard Model).  Specifically,
we examine the contributions to $S$, $T$ and $U$ from:
\item{A.} The squarks and sleptons.  Note that each squark and each
slepton generation (\ie, summing over both up-type and
down-type superpartners)
contributes separately a finite result to $S$, $T$
and $U$.
\item{B.} The neutralinos and charginos.
\item{C.} The MSSM Higgs sector.

\noindent
The treatment of the MSSM Higgs sector requires some care to
insure a finite contribution to $S$, $T$ and $U$.  As before,
one must first
define the RSM.  Then,
$$\eqalign{S&=S\ls{\rm RSM}+\delta S\,,\cr
T&=T\ls{\rm RSM}+\delta T\,,\cr
U&=U\ls{\rm RSM}+\delta U\,,\cr}
\eqn\stsusy$$
where the MSSM Higgs sector contributions to
$\delta S,\ \delta T$ and $\delta U$
are obtained from eq.~\studefs\
by computing the MSSM Higgs loops contributing to $A_{ij}(0)$ and
$F_{ij}$ (including diagrams
with one virtual Higgs boson and one virtual gauge boson) and
subtracting off the corresponding Higgs loops of the RSM.
In the present case, it is convenient to
define the RSM to be the Standard Model
with the Standard Model Higgs boson mass set
equal to the mass of the lightest CP-even Higgs scalar of the MSSM.
In addition, until $\mt$ is known, the definitions of $\delta S$,
$\delta T$ and $\delta U$ will depend on the value of $\mt$ chosen
for the RSM. Typically, one chooses $\mt=\mz$ (equal to the present
experimental CDF lower bound\refmark\cdflimit)
in order to obtain conservative limits on the possible new physics
contributions to $S$, $T$  and $U$.

Some of the results for the MSSM contributions to $S$, $T$ and
$U$ are presented below.  The qualitative behavior of these
results is easily summarized.  The contribution of any given
sector to $\delta S$, $\delta T$ and $\delta U$ behaves as
$$\delta S({\rm MSSM})\sim\delta T({\rm MSSM})\sim
{\cal O}\left({\mz^2\over\msusy^2}\right)\,,\eqn\susyst$$
$$\delta U({\rm MSSM})\sim
{\cal O}\left({\mz^4\over\msusy^4}\right)\,,\eqn\susyu$$
in the limit where $\msusy\gg\mz$.  In each sector, $\msusy$
corresponds roughly to the mass parameter that contributes the
dominant part of the corresponding sector masses.\foot{These
masses are: in sector A, the SU(2)$\times$U(1) conserving
diagonal squark and slepton masses ($M_{\widetilde Q}$, $M_{\widetilde U}$,
$M_{\widetilde D}$, $M_{\widetilde L}$ and $M_{\widetilde E}$) of
eq.~\diagsqmasses;
in sector B, the gaugino mass parameters $M_1$ and $M_2$ and the
supersymmetric Higgs mass parameter $\mu$; and in sector C, the
CP-odd Higgs mass, $\mha$.}
Thus, all MSSM contributions
to $S$, $T$ and $U$ vanish at least quadratically in $\msusy$ as the
supersymmetric particle masses become large.  Unlike the
non-supersymmetric examples presented previously,
the effects of the supersymmetric particles (and all Higgs bosons
beyond $\hl$) smoothly decouple; the resulting low-energy effective
theory at the scale $m_Z$ is precisely that of the Standard Model.
This decoupling behavior is easily understood.
As an example, consider the effects of the MSSM Higgs sector.
As indicated above, the sum of
the contributions to $S$, $T$ and $U$ is finite after
subtracting off the contribution of the Standard Model
Higgs boson with $\mhsm=\mhl$.  According to the results
of section 2, the mass of $\hl$ cannot be arbitrarily large---it is
bounded at tree level by $\mz$.  All other Higgs masses can become
large by taking $\mha\gg\mz$.  In this limit, we see
that $\mhpm\simeq\mhh\simeq\mha$ and $\mhl\simeq\mz|\cos2\beta|$.
However, in this limit, the large Higgs masses are due  to the
large value of the mass parameter $m_{12}$ [see eq.~\susymhpm]
rather than a large Higgs self-coupling (which is the case in the
large Higgs mass limit of the Standard Model).  In particular, the
Higgs self-couplings in the MSSM are gauge couplings which can
never become large.  As a result, the decoupling theorem applies.
Similar arguments can be applied to the other sectors.  Supersymmetric
particle masses can be taken large by increasing the values of
SU(2)$\times$U(1) conserving mass parameters.  Thus, the virtual effects
of heavy supersymmetric particles must decouple.\foot{One can also
show that the coupling of the supersymmetric fermions (charginos
and neutralinos) to the massive gauge bosons become purely
vector-like in the limit of large fermion mass.  Thus, again the
decoupling theorem applies to the heavy virtual supersymmetric
fermion exchanges.}

Let us now turn to some specific calculations.  The only remaining
question is to evaluate the constant of proportionality that
is implicit in eqs.~\susyst\ and \susyu.  In particular, note
that $\delta U\ll \delta S$, $\delta T$,
so I shall focus primarily
on $\delta S$ and $\delta T$ below.  It is important
to emphasize that the $S$, $T$, $U$ formalism is useful only in
the limit where the new physics is sufficiently
heavy as compared to $\mz$.  In the present application,
``sufficiently heavy'' means that it is a good
approximation to keep only the leading
${\cal O}(\mz^2/\msusy^2)$ terms.  Because of various additional
factors such as $1/16\pi^2$
which typically arise in loop-calculations,
I expect the results presented
below to be reasonably accurate for supersymmetric masses above, say,
150 GeV.  For lighter supersymmetric particle masses,
more precise computations are required.\foot{However,
note that the calculation of $T$ (or
the $\rho$ parameter) is meaningful over the entire range of possible
supersymmetric masses.}

Consider first the contributions
of the top/bottom squark system.  From eqs.~\diagsqmasses--\mlr,
one can easily diagonalize the
two $2\times 2$ squark mass matrices to find the mass eigenstates
and corresponding mixing angles.  To a good approximation, we can
neglect the mixing between $\widetilde b_L$ and $\widetilde b_R$.
Denote the top-squark states $\widetilde t_1$ and $\widetilde t_2$
with corresponding mixing angle $\theta_t$.  Then, in terms of
the function $F$ defined in eq.~\effdef,
$$\eqalign{
\delta\rho(\widetilde t,\widetilde b)=&{g^2 N_c\over 32\pi^2\mw^2}
\left[\cos^2\theta_t F(m^2_{\tilde t_1},m^2_{\tilde b_L})+
\sin^2\theta_t F(m^2_{\tilde t_2},m^2_{\tilde b_L})\right.
\cr &\qquad\qquad\quad\left.-
\sin^2\theta_t\cos^2\theta_t F(m^2_{\tilde t_1},m^2_{\tilde t_2})
\right]\,.\cr}\eqn\deltarhosq$$
$F(m_1^2,m_2^2)$ has the following properties: in the
limit of $|m_1^2-m_2^2|\ll m_1^2$, $m_2^2$,
$$\eqalign{%
F(m_1^2,m_2^2) &\simeq {(m_1^2-m_2^2)^2\over 6m_2^2}\ , \crr
F(m_1^2,m_3^2)-F(m_2^2,m_3^2) &\simeq (m_1^2-m_2^2)
  \left[{1\over 2}+ {m_3^2 \over m^2_3-m^2_2}
  + {m^4_3\over (m^2_3-m^2_2)^2} \ln\left({m^2_2\over m^2_3}\right)
  \right]\, . \cr}
\eqn\flimit$$
Using these results and explicit formulae for the top-squark masses
and mixing angles in terms of the parameters given in eqs.~\diagsqmasses\
and \mlr, one ends up with
$$\delta\rho(\widetilde t,\widetilde b)\simeq  {g^2N_c m^4_t C\over
   32\pi^2m^2_W \msusy^2}
\,,\eqn\rhostsb$$
where I have assumed that  $\msusy \gg \mz,\mt$. Here, $\msusy$
is the largest of the supersymmetric mass parameters that appear
in the squark mass matrix, and $C$ is a dimensionless function
of the model parameters.  In particular, $C$ approaches a constant
in the limit where one or more of the various supersymmetric
masses become large, so that eq.~\rhostsb\ exhibits the
expected decoupling behavior.  It may appear that the
correction exhibited in eq.~\rhostsb\ is significantly enhanced by
a factor of $\mt^4$.  However, this result is deceptive, since we
must compare it to the corresponding Standard Model result
[eq.~\rhocorrection].  One sees that the MSSM contribution is
in fact suppressed by $\mt^2/\msusy^2$ relative to the Standard Model
result.  Of course, this is not a suppression, if the relevant
supersymmetric mass parameters are small.  For example,
in the supersymmetric limit ($M_{\tilde t_1}=M_{\tilde t_2}=\mt$ and
$M_{\tilde b_L}=\mb$), one finds $\delta\rho(\widetilde t,\widetilde b)=
\delta\rho(t,b)$, in which case $\delta\rho$ would be
twice as large as its predicted Standard Model value.  This
demonstrates that an appreciable MSSM contribution to precision
electroweak measurements is possible in principle if supersymmetric
particle masses are light enough.  However, once these masses are taken
larger than $\mt$, their contributions to one-loop effects diminish
rapidly.

%\topinsert
%   \tenpoint \baselineskip=12pt   \narrower
%\plotpicture{\hsize}{8cm}{stsqk.topdraw}
%\vskip12pt\noindent
%{\bf Fig.~\stsqk.}\enskip
%\endinsert
\FIG\stsqk{%
The contribution to the $S$ and $T$ parameters from
the squark and slepton sector of the MSSM as a function
of $M_{\widetilde Q}$.  The squark mass spectrum is determined
by MSSM mass parameters defined in eqs.~\diagsqmasses\ and \mlr.
For simplicity, all soft supersymmetry breaking diagonal
squark (and slepton) mass parameters are taken
equal to $M_{\widetilde Q}$.
In addition, $A=M_{\widetilde Q}$, $\mu=-200$~GeV and
$\tan\beta=2$
are the parameters that determine the strength of the
off-diagonal squark mixing.  The solid curves show the contributions
of three generations of squarks and sleptons, while the dashed curves
show the partial contribution arising from the top/bottom squark
sector alone.}
\FIG\stchi{%
The contribution to the $S$ and $T$ parameters from the
neutralino and chargino sector of the MSSM as a function
of $\mu$ for $\tan\beta=2$.  The four curves shown correspond
to $M=50$, $250$, $500$ and $1000$~GeV [with $M_2\equiv M$ and
$M_1=(5g^{\prime 2}/3g^2)M$].  In (a), curves in the region
of $|\mu|\leq100$~GeV are not shown, since in this region of
parameter space the light chargino mass is of order $m_Z$ (or less).
In (b), $T$ is related to the
$\rho$ parameter via $\delta\rho=\alpha\delta T$
which is an experimental observable over the entire mass parameter
region.}

In fig.~\stsqk, I plot the squark and slepton contributions to
$S$ and $T$ for a typical set of MSSM parameters.  For simplicity,
all diagonal squark soft-supersymmetry breaking parameters
have been taken equal to $M_{\widetilde Q}$.  In addition, I have
included squark mixing (which is appreciable only in the top
squark sector) by taking all $A$-parameters equal to $M_{\widetilde Q}$.
For illustrative purposes, I have chosen
$\mu=-200$~GeV (and $\tan\beta=2$) which enhances somewhat the top
squark mixing.  This choice of parameters leads to a large mass splitting
of the top squark eigenstates (and a rather light top squark) when
$M_{\widetilde Q}$ is near its lower limit as shown in fig.~\stsqk.
Nevertheless, we see that the contributions to $S$ and $T$ from the
squark and slepton sector never exceed 0.1 for the parameters shown.
As suggested above, larger values for $\delta T$ can occur only
for parameter choices approaching the supersymmetric limit.

Consider next the contributions to $S$ and $T$ from the neutralino
and chargino sector, shown in fig.~\stchi.  Again, one never sees
values of $\delta S$ and $\delta T$ larger than 0.1.  I have omitted
plotting $\delta S$ in the region of small $\mu$, since the lightest
chargino mass is $\lsim\mz$ in this region of parameter space.
In this case, it is no
longer true that all particle masses associated with the ``new physics''
lie significantly above $\mz$, so that the assumption that the
radiative corrections are simply parametrized by $S$, $T$ and $U$ breaks
down.

%topinsert
%  \tenpoint \baselineskip=12pt   \narrower
%plotpicture{\hsize}{8cm}{stchi.topdraw}
%vskip12pt\noindent
%{\bf Fig.~\stchi.}\enskip
%\vskip15pt
%\endinsert

\FIG\sthiggs{%
The contribution to the $S$ and $T$ parameters from
the MSSM Higgs sector
relative to the Standard Model with Higgs mass set equal to $\mhl$,
as a function of $\mha$.  Three curves corresponding to
$\tanb= 1$, 2, and 10 are shown.  In (a), the dotted curve corresponds
to the asymptotic prediction [eq.~\shiggsapprox] for $\tanb=10$.
In the region of small $\mha$ where the dotted curve departs from the
dashed curve, it is no longer useful to use $S$ in the
parametrization of the radiative
corrections.  In (b), the two dotted curves correspond to
the asymptotic predictions [eq.~\thiggsapprox]
for $\tanb=1$ and 10 respectively.}
%Note that $T$ is related to the
%$\rho$ parameter via $\delta\rho=\alpha\delta T$
%which is an experimental observable for all values of $\mha$.

Finally, I consider the contributions to $S$ and $T$ from the MSSM
Higgs sector.  Here one must be careful to define these contributions
relative to the RSM, as discussed above.  In this case, I subtract
out the contribution of the Standard Model Higgs boson whose mass
is chosen to be equal to the mass of the lightest CP-even scalar
($\mhl$).  The results of an exact one-loop
computation of the MSSM Higgs
contributions to $S$, $T$ and $U$ are given in Appendix
C\refmark\hehth.  (See refs.~[\doug,\higgsrad--\higgsradtwo] for
previous work on radiative corrections in two-Higgs doublet models.)
Numerical
results are shown in fig.~\sthiggs.  To understand why
the numerical values for the MSSM-Higgs contributions to
$\delta S$ and $\delta T$ are particularly small, it is
instructive to evaluate the corresponding
expressions of Appendix C in the
limit of large $\mha$.  I find\foot{Asymptotically, $\delta U$%
(MSSM--Higgs) = ${\cal O}(m^4_Z/m^4_A)$, which is completely
negligible.}
$$\delta S({\rm MSSM}\!-\!{\rm Higgs})
\simeq{\mz^2(\sin^2 2\beta-2\cos^2\theta_W)\over 24\pi\mha^2}\,,
\eqn\shiggsapprox$$
$$\delta T({\rm MSSM}\!-\!{\rm Higgs})
\simeq{\mz^2(\cos^2\theta_W-\sin^2 2\beta)\over 48\pi\mha^2
\sin^2\theta_W}\,.\eqn\thiggsapprox$$

%\topinsert
%   \tenpoint \baselineskip=12pt  \narrower
%\plotpicture{\hsize}{8cm}{sthiggs.topdraw}
%\vskip6pt\noindent
%{\bf Fig.~\sthiggs.}\enskip
%
%\endinsert

An analysis of $S$, $T$ and $U$ based on LEP data (assuming a
RSM where $\mt=\mhsm=\mz$) reported in ref.~[\paul] yields:
$\delta S=-0.97\pm 0.67$, $\delta T=-0.18\pm 0.51$ and
$\delta U=0.07\pm 0.97$.
It is hard to imagine that the these quantities
could ever be measured to an accuracy better than 0.1, whereas
the analysis above implies that the contributions of the MSSM to
$S$ and $T$ must lie below 0.1 if all supersymmetric particle masses
are above $\mz$.
One must also consider the possibility of other contributions
to $\delta S$ and $\delta T$.  As long as $\mt$ is not well known,
there will be $\mt$ dependence in these quantities (entering through the
$\mt$ choice of the RSM).  However,
even when $\mt$ is known with some accuracy, it is doubtful
that virtual effects of the MSSM can ever be detected
via its oblique radiative corrections.

\bigskip
\leftline{{\bf 4. Radiative Corrections to Processes Involving
$\bf b$ (and $\bf t$) Quarks}}
\medskip

\REF\gapjf{G. Altarelli and P. Franzini, {\sl Z.~Phys.} {\bf C37}
(1988) 271; G.G. Athanasiu, P.J. Franzini and F.J. Gilman,
{\sl Phys. Rev.} {\bf D32} (1985) 3010;
S.L. Glashow and E.E. Jenkins, {\sl Phys. Lett.} {\bf B196} (1987) 233;
F. Hoogeveen and C.N. Leung, {\sl Phys. Rev.} {\bf D37} (1988) 3340;
J.F. Gunion and B. Grzadkowski, {\sl Phys. Lett.} {\bf B243}
(1990) 301.}
\REF\cqgjnn{C.Q. Geng and J.N. Ng, {\sl Phys. Rev.} {\bf D38}
(1988) 2857.}
\REF\stefan{
S. Bertolini, F. Borzumati, A. Masiero and G. Ridolfi,
{\sl Nucl. Phys.} {\bf B353} (1991) 591.}
\REF\buras{A. Buras, P. Krawczyk, M.E. Lautenbacher and C. Salazar,
{\sl Nucl. Phys.} {\bf B337} (1990) 284.}
\REF\hewett{V. Barger, J.L. Hewett and R.J.N. Phillips, {\sl Phys.
Rev.} {\bf D41} (1990) 3421.}
\REF\zbbar{A. Djouadi, G. Girardi, C. Verzegnassi, W. Hollik and
F.M. Renard, {\sl Nucl. Phys.} {\bf B349} (1991) 48; M. Boulware and
D. Finnell, {\sl Phys. Rev.} {\bf D44} (1991) 2054.}
\REF\wise{B. Grinstein and
M.B. Wise, {\sl Phys. Lett.} {\bf B201} (1988) 274.}
\REF\willey{W.-S. Hou and R.S. Willey, {\sl Phys. Lett.} {\bf B202}
(1988) 591; {\sl Nucl. Phys.} {\bf B326} (1989) 54;
T. Rizzo, {\sl Phys. Rev.} {\bf D38} (1988) 820;
X.-G. He, T.D. Nguyen and R.R. Volkas, {\sl Phys. Rev.} {\bf D38}
(1988) 814; M. Ciuchini, {\sl Mod. Phys. Lett.} {\bf A4}
(1989) 1945.}
\REF\joanne{J.L. Hewett, {\sl Phys. Rev. Lett.} {\bf 70} (1993) 1045;
V. Barger, M.S. Berger, and R.J.N. Phillips, {\sl Phys. Rev. Lett.}
{\bf 70} (1993) 1368.}
\REF\joannetwo{J.L. Hewett, Argonne preprint ANL-HEP-PR-93-21 (1993).}
\REF\barbii{R. Barbieri and G.F. Giudice, CERN-TH.6830/93 (1993).}

In section 3, I demonstrated that oblique radiative corrections
due to supersymmetric particle exchange are not observable if
all supersymmetric particle masses lie above $\mz$.  However,
this leaves the possibility that certain non-oblique radiative
corrections might be detectable.  In this section, I shall briefly
address this possibility.

A promising class of non-oblique radiative corrections consists
of processes that involve an external $b$-quark.
%For example,
%vertex corrections and box diagrams
%that involve an intermediate $t$-quark
%and charged Higgs boson can be substantial, because $g_{H^-t\bar b}$
%contains a piece proportional to $m_t\cot\beta/m_W$ [see eq.~\hpmqq].
Three examples of such processes that have been studied in the
literature are: (i) charged Higgs box diagram
contributions to $B^0$--$\overline{B^0}$
mixing\refmark{\gapjf--\hewett}; (ii) the charged Higgs
vertex correction to $Z\to b\bar b$\refmark\zbbar;
and (iii) the charged Higgs vertex corrections to rare
$b$-decays\refmark{\cqgjnn--\buras,\wise--\joannetwo}\
such as $b\to s\gamma$,
$b\to s\ell^+\ell^-$, $b\to sg$ and $b\to s\nu\bar\nu$.
Of course, in the MSSM
there will also be supersymmetric particle contributions to
all of the one-loop processes mentioned above.
Some of these contributions
(\eg, loops containing top-squarks) could be as important as
the charged Higgs effects.  Depending on the sign of the relative
contributions (which depends in detail on the MSSM parameters),
the overall MSSM contribution to the various rare $b$ decays could
be substantially different from the
charged Higgs effects alone\refmark{\stefan,\zbbar,\barbii}.

Among the rare $b$-decays, the charged Higgs contribution to
$b\to s\gamma$ is perhaps the most promising.  The theoretical
prediction for this rate in the Standard Model is
$BR(B\to K\gamma+X)\simeq 2.8\times 10^{-4}$ ($3.4\times 10^{-4}$),
for $\mt=150$ (200) GeV, where both the leading log and the first
non-leading log QCD corrections have been
included\refmark\joannetwo.  Incorporating the
charged Higgs contribution [assuming an
$H^-t\bar b$ coupling as specified in eq.~\hpmqq] enhances the $b\to s
\gamma$ branching ratio over the Standard Model expectation.  An
explicit computation shows that the amplitude for $b\to s\gamma$
in the two-Higgs doublet model (omitting supersymmetric particle
contributions) has the following structure
$$\eqalign{%
{\cal M}(b\to s\gamma) &={eg^2m_b^3 V_{tb} V^\ast_{ts}\over\mw^2}\crr
&\quad \times\left[ A_W+(A_{H_1}\cot^2\beta+A_{H_2}) \left.
\times\cases{1,&$\mhpm\ll\mt$\cr
\mt^2/\mhpm^2\,,&$\mt\ll\mhpm$\cr} \right\}\right] \cr}
\eqn\bsgam$$
where $A_W$, $A_{H_1}$ and $A_{H_2}$ are dimensionless functions of
the particle masses arising from the loop graphs involving $W$ and
$\hpm$ exchange.  In particular, $A_{H_1}$ and $A_{H_2}$ approach
finite non-zero constants in the two limiting cases indicated above.
Eq.~\bsgam\ illustrates two of the features encountered in section 3.
On one hand, ones sees the non-decoupling of the top-quark in the
large $\mt$ limit.  In this case, the amplitude approaches an
$\mt$-independent constant when $\mt$ is larger than all mass scales
in the problem.  On the other hand, when $\mhpm\gg\mt$, the effect
of the charged Higgs loop decouples quadratically with the Higgs mass.
This behavior
is expected for the same reasons that heavy supersymmetric particles
decouple from oblique radiative corrections in the large $\msusy$ limit.

%\topinsert
%   \tenpoint \baselineskip=12pt  \narrower
%\plotpicture{\hsize}{8cm}{bsgamen.topdraw}
%\vskip6pt\noindent
%{\bf Fig.~\bsgamen.}\enskip
\FIG\bsgamen{%
The ratio of $BR(B\to K\gamma+X)$ in the two-Higgs-doublet
model relative to its predicted value
in the Standard Model (SM) as a function of the charged
Higgs mass for $\mt=150$ and 200~GeV and various choices for $\tanb$,
assuming an $H^-t\bar b$ coupling given by eq.~\hpmqq.
This graph is based on calculations of ref.~[\buras].}
%\vskip15pt
%\endinsert

\REF\cleolim{D.L. Kreinick, CBX 92-85 (1992); S. Sanghera \etal\
[CLEO Collaboration], contributed paper to the Dallas Conference (1992).}
All that is left to do is to explore the relative sizes
of the coefficients in eq.~\bsgam\ and to incorporate the (leading log)
QCD radiative corrections\refmark\wise\
in order to see how fast the decoupling of a heavy $\hpm$ occurs.
%It is at this point where we
%encounter a surprise (particularly in light of the numerical
%results presented in section 3).
Using the explicit formulae
in ref.~[\buras], I have plotted in fig.~\bsgamen\
the enhancement of the branching ratio
for $b\to s\gamma$ in the two-Higgs-doublet model relative to the
Standard Model rate.  It follows from
eq.~\bsgam\ that the decoupling of the
$\hpm$ contribution is controlled by the factor $m_t^2/\mhpm^2$
[in contrast to $\mz^2/\mhpm^2$ as in eqs.~\shiggsapprox-\thiggsapprox].
Even so, it is perhaps surprising that the significance of the
charged Higgs exchange to $b\to s\gamma$ persists to such large values
of $\mhpm$.
In addition, the $\tan\beta$ dependence of the curves in fig.~\bsgamen\
reflect the contribution in eq.~\bsgam\ proportional to
the square of the Higgs-top quark Yukawa coupling, $\lambda_t^2$.
(The $\tan\beta$-independent part of the charged Higgs contribution
in eq.~\bsgam\ arises from a term proportional to $\lambda_t\lambda_b$.)
The current experimental limit from the CLEO
Collaboration\refmark\cleolim\
of $BR(B\to K\gamma+X)<8.4\times 10^{-4}$ (at $90\%$ CL)
already places
interesting limits on the parameters of the charged Higgs sector
(see refs.~[\joanne] and [\joannetwo]
for a recent analysis of these constraints).
Forthcoming improved limits from CLEO (or an observed signal)
could significantly constrain $\mhpm$ and $\tanb$ and may lead to other
important restrictions on supersymmetric particle masses.

\bigskip\bigskip
\leftline{{\bf 5. Radiative Corrections to the MSSM Higgs Sector}}
\medskip

\REF\happroxz{Z. Kunszt and W.J. Stirling, \sl Phys. Lett. \bf B242 \rm
(1990) 507; N. Brown, \sl Z. Phys. \bf C49 \rm (1991) 657; V. Barger,
R.J.N. Phillips and K. Whisnant, {\sl Phys. Rev.} {\bf D43} (1991) 1110;
D.J. Summers, {\sl Phys. Lett.} {\bf B274} (1992) 209.}
\REF\hhprl{H.E. Haber and R. Hempfling, {\sl Phys. Rev. Lett.} {\bf 66}
(1991) 1815.}
\REF\radmssm{Y.
Okada, M. Yamaguchi and T. Yanagida, {\sl Prog. Theor. Phys.} {\bf 85}
(1991) 1; {\sl Phys. Lett.} {\bf B262} (1991) 54;
J. Ellis, G. Ridolfi and F. Zwirner, {\sl Phys. Lett.}
{\bf B257} (1991) 83; {\bf B262} (1991) 477; R. Barbieri, M. Frigeni,
and F. Caravaglios, {\sl Phys. Lett.} {\bf B258} (1991) 167.}
\REF\moreradmssm{%
R. Barbieri and M. Frigeni, {\sl Phys. Lett.} {\bf B258} (1991) 395;
A. Yamada, {\sl Phys. Lett.} {\bf B263} (1991) 233.}
\REF\quiros{J.R. Espinosa and M. Quiros, {\sl Phys. Lett.} {\bf B267}
(1991) 27.}
\REF\pokorski{
P.H. Chankowski, S. Pokorski and J. Rosiek, {\sl Phys. Lett.}
{\bf B274} (1992) 191; {\bf B281} (1992) 100.}
\REF\llog{H.E. Haber and R. Hempfling, SCIPP 91/33 (1992).}
\REF\berkeley{D.M. Pierce, A. Papadopoulos, and S. Johnson,
{\sl Phys. Rev. Lett.} {\bf 68} (1992) 3678.}
\REF\sasaki{K. Sasaki, M. Carena and C.E.M. Wagner, {\sl Nucl. Phys.}
{\bf B381} (1992) 66.}
\REF\andrea{A. Brignole, {\sl Phys. Lett.} {\bf B281} (1992) 284.}
\REF\diaztwo{M.A. Diaz and H.E. Haber, {\sl Phys.
Rev.} {\bf D46} (1992) 3086.}
The tree-level Higgs mass
predictions of section 2 have important phenomenological
consequences.  For
example, the bound $\mhl\leq\mz$, if reliable, would have significant
implications for future experiments
at LEP-II.  In principle, experiments running at LEP-II operating
at $\sqrt{s}=200$~GeV and design luminosity would either discover the
Higgs boson (via $e^+e^-\to\hl Z$)
or rule out the MSSM.  (Whether this is possible to do in practice
depends on whether Higgs bosons with $\mhl\approx\mz$ can be detected%
\refmark\happroxz.)  However, the tree-level Higgs mass relations
are examples of {\it natural relations} of the theory. Such relations
suffer finite (and therefore calculable) radiative corrections.  In
particular, $\mhl\leq\mz$ need not be respected
when radiative corrections are incorporated.  For example,
in the radiative corrections to the neutral CP-even Higgs squared-mass
matrix, the $22$-element is shifted by a term proportional to
$(g^2 m_t^4/\mw^2)\,\ln(M_{\tilde t}^2/m_t^2)$%
\refmark{\hhprl--\diaztwo}.
Such a term arises from an incomplete cancellation
between top-quark and top-squark
loop contributions to the neutral Higgs boson self-energy.
If $m_t$ is large, this term significantly alters the tree-level
predictions.  In this section, I will compute the radiative
corrections to MSSM Higgs boson masses in the leading logarithmic
approximation.  This approximation scheme is simple to carry out
and neatly summarizes the main correction terms.
\REF\hemperice{R. Hempfling, contribution to These Proceedings.}
(For a discussion of the impact of the non-leading logarithmic
corrections, see ref.~[\hemperice].)
%%%%%%%%%%%%%%%%%%%%%%%%%%%%%%%%%%%%%%%%%%%%%%%%%%%%%%%%%%%%%%%%%%%%%%%

\def\crrr{\cr\noalign{\vskip8pt}}

The complete one-loop computation of the MSSM Higgs masses can be
found in the literature\refmark{\pokorski,\andrea}.  However,
the formulae involved are very lengthy and not too transparent.
Instead, I will present here
the results based on a calculation of the Higgs
mass-squared matrix in which all leading logarithmic terms are
included (see ref.~[\llog] for details).
We take the supersymmetry breaking
scale ($\msusy$) to be somewhat larger than the electroweak scale.
For simplicity, we assume that the masses of all supersymmetric
particles (squarks, sleptons, neutralinos and charginos) are roughly
degenerate and of order $\msusy$.
Admittedly, this is a crude approximation.
However, deviations from this assumption will
lead to non-leading logarithmic corrections which tend to be small
if the supersymmetric particles are not widely split in mass.
Moreover, the procedure outlined below can be modified to
incorporate the largest non-leading logarithmic contributions that
arise in the case of multiple supersymmetric particle thresholds and/or
large squark mixing\refmark{\llog,\hemperice}.

The leading logarithmic expressions for the MSSM
Higgs masses are obtained from
eqs.~\mamthree\ and \massmhh\ by treating the $\lambda_i$ as
running parameters evaluated at the electroweak scale, $\mweak$.
In addition, we identify the $W$ and $Z$ masses by
$$\eqalign{\mw^2&=\fourth g^2(v_1^2+v_2^2)\,,\crr
\mz^2&=\fourth (g^2+g'^2)(v_1^2+v_2^2)\,,\cr}\eqn\vmasses$$
where the running gauge couplings are also evaluated at $\mweak$.
Of course, the gauge couplings, $g$ and $g'$ are known from
experimental measurements which are performed at the scale $\mweak$.
The $\lambda_i(\mweak^2)$ are determined from supersymmetry.
Namely, if supersymmetry were unbroken, then the $\lambda_i$ would
be fixed according to eq.~\bndfr.  Since supersymmetry is broken,
we regard eq.~\bndfr\ as boundary conditions for the running
parameters, valid at (and above) the energy scale $\msusy$.  That is,
we take
$$\eqalign{
\lambda_1(\msusy^2)&=\lambda_2(\msusy^2)=\fourth\left[g^2(\msusy^2)
+g'^2(\msusy^2)\right],\crrr
\lambda_3(\msusy^2)&=\fourth\left[g^2(\msusy^2)-g'^2(\msusy^2)\right],\crrr
\lambda_4(\msusy^2)&=-\half g^2(\msusy^2),\crrr
\lambda_5(\msusy^2)&=\lambda_6(\msusy^2)=
\lambda_7(\msusy^2)=0\,,\cr}\eqn\boundary$$
%\vskip5pt
in accordance with the tree-level relations of the MSSM.  At scales
below $\msusy$, the gauge and
quartic couplings evolve according to the
renormalization group equations (RGEs) of the non-supersymmetric
two-Higgs-doublet model given in eqs.~\betagg--\betal.
These equations are of the form:
\vskip5pt
$${dp_i\over dt} =
\beta_i(p_1,p_2,\ldots)\qquad\hbox{with}~t\equiv\ln\,\mu^2
\,,\eqn\rgeqs$$
\vskip5pt\noindent
where $\mu$ is the energy scale, and
the $p_i$ are the
parameters of the theory ($p_i = g_j^2,\lambda_k,\ldots$).
The relevant $\beta$-functions can be found in Appendix B.
The boundary conditions together with the RGEs imply that, at the
leading-log level, $\lambda_5$, $\lambda_6$ and $\lambda_7$ are zero
at all energy scales.  Solving the RGEs with
the supersymmetric boundary conditions at $\msusy$, one can determine
the $\lambda_i$ at the weak scale.
%\foot{In doing so, one must
%also run the gauge couplings that appear in eq.~\boundary\ down from
%$\msusy$ to $\mweak$.}
The resulting values for $\lambda_i(\mweak)$
are then inserted into eqs.~\mamthree\ and \massmhh\ to obtain the
radiatively corrected Higgs masses.  Having solved the one-loop RGEs,
the Higgs masses thus obtained  include the leading
logarithmic radiative corrections summed to all orders in perturbation
theory.

The RGEs can be solved by numerical analysis on the computer.  But
it is instructive to solve the RGEs iteratively.  In first
approximation, we can take the right hand side of eq.~\rgeqs\ to
be independent of $\mu^2$.  That is, we compute the $\beta_i$ by
evaluating the parameters $p_i$ at the scale $\mu=\msusy$.
Then, integration of the RGEs is
trivial, and we obtain
\vskip5pt
$$p_i(\mweak^2)=p_i(\msusy^2)-\beta_i\,\ln\left({\msusy^2\over\mweak^2}
\right)\,.\eqn\oneloopllog$$
\vskip5pt\noindent
Note that this iterative solution corresponds to computing the
one-loop radiative corrections in which only terms proportional to
$\ln\msusy^2$ are kept.  It is straightforward to work out the
one-loop leading logarithmic expressions for the $\lambda_i$ and
the Higgs masses.  First consider the charged Higgs mass.  Since
$\lambda_5(\mu^2)=0$ at all scales, we need only consider $\lambda_4$.
Evaluating $\beta_{\lambda_4}$ at $\mu=\msusy$, we compute
\vskip5pt
$$\eqalign{%
\lambda_4(\mw^2)=-\half g^2
-&{1\over{32\pi^2}}\biggl[\bigl({\textstyle{
4\over 3}}N_g+{\textstyle{1\over 6}}N_H-{\textstyle{10\over 3}}
\bigr)g^4+5g^2g'^2\crr
&\qquad\qquad-{{3g^4}\over{2m_W^2}}\left({{m_t^2}\over
{s_{\beta}^2}}+{{m_b^2}\over{c_{\beta}^2}}\right)
+{{3g^2m_t^2m_b^2}\over{s_{\beta}^2c_{\beta}^2m_W^4}}\Biggr]
\ln{{\msusy^2}\over{\mw^2}}\,,\cr}\eqn\lcuaunloop$$
\vskip5pt\noindent
where $s_\beta\equiv\sin\beta$ and $c_\beta\equiv\cos\beta$.
The terms proportional to the number of generations $N_g=3$
and the number of Higgs doublets $N_H=2$ that remain in the
low-energy effective theory at the scale $\mu=\mw$ have their origin in
the running of $g^2$ from $\msusy$ down to $\mw$.
In deriving this expression, I have taken $\mweak=\mw$.  This is
a somewhat arbitrary decision, since another reasonable choice would
yield a result that differs from eq.~\lcuaunloop\ by a non-leading
logarithmic term.  Comparisons with a more complete
calculation show that
one should choose $\mweak=\mw$ in computations involving the charged
Higgs (and gauge) sector and $\mweak=\mz$ in computations involving the
neutral sector.

\REF\diaz{M.A. Diaz and H.E. Haber, {\sl Phys. Rev.} {\bf D45} (1992)
4246.}
The above analysis also assumes that $m_t\sim {\cal O}(m_W)$.  Since
$\mt>\mw$, one can improve the above result somewhat
by decoupling the $(t,b)$ weak doublet from
the low-energy theory for scales below $m_t$.  The
terms in eq.~\lcuaunloop\ that are proportional to $m_t^2$ and/or
$m_b^2$ arise from self-energy diagrams containing a $tb$ loop.
Thus, such a term should not be present for $\mw\leq \mu\leq m_t$.
In addition, we recognize the term in eq.~\lcuaunloop\ proportional
to the number of generations $N_g$ as arising from the contributions
to the self-energy diagrams containing either quark or lepton loops
(and their supersymmetric partners).
To identify the contribution of the $tb$ loop to this term,
simply
write
\vskip3pt
$$N_g=\fourth N_g(N_c+1)=\fourth N_c+\fourth[N_c(N_g-1)+N_g]\,,
\eqn\ngee$$
\vskip5pt\noindent
where $N_c=3$ colors.  Thus, we identify ${1\over 4}N_c$ as the piece
of the term proportional to $N_g$ that is due to the $tb$ loop.  The
rest of this term is then attributed to the lighter quarks and leptons.
Finally, the remaining terms in eq.~\lcuaunloop\ are due to the
contributions from the gauge and
Higgs boson sector.  The final result is\refmark\diaz\
\vskip5pt
$$\eqalign{\lambda_4(\mw^2) &=-\half g^2
-{N_c g^4\over{32\pi^2}}\left[{1\over 3}
-{1\over{2m_W^2}}\biggl({{m_t^2}\over
{s_{\beta}^2}}+{{m_b^2}\over{c_{\beta}^2}}\biggr)
+{{m_t^2m_b^2}\over{s_{\beta}^2c_{\beta}^2m_W^4}}\right]
\ln{{\msusy^2}\over{m_t^2}}\crr
&\qquad-{1\over 96\pi^2}\left\{\left[N_c(N_g-1)+N_g
+\half N_H-10\right]g^4
+15g^2g'^2 \right\}\ln{\msusy^2\over\mw^2}
\,.\cr}\eqn\lambdaiv$$
\vskip12pt\noindent
Inserting this result (and $\lambda_5=0$) into eq.~\mamthree,
we obtain the one-loop leading-log formula for the charged Higgs mass
\vskip3pt
$$\eqalign{%
m_{H^{\pm}}^2&=m_A^2+m_W^2 +{{N_c g^2}\over{32\pi^2m_W^2}}
\Bigg[{{2m_t^2m_b^2}\over{s_{\beta}^2c_{\beta}^2}}-m_W^2
\bigg({{m_t^2}\over{s_{\beta}^2}}+{{m_b^2}\over{c_{\beta}^2}}\bigg)
+{\textstyle{2\over 3}}m_W^4\Bigg]\ln{{\msusy^2}\over{m_t^2}}  \crr\crr
&\qquad+{{m_W^2}\over{48\pi^2}} \Big\{\left[N_c(N_g-1)+N_g
+\half N_H-10\right]g^2 +15g'^2\Big\}
\ln{{\msusy^2}\over{m_W^2}}\,.
\cr}\eqn\llform$$
\vskip5pt\noindent
Since this derivation makes use of the two-Higgs-doublet RGEs for
the $\lambda_i$, there is an implicit assumption that the full
two-doublet Higgs spectrum survives in the low-energy effective theory
at $\mu=\mw$.  This means that we {\it must} take $N_H=2$ in the
formulae above.  It also means that $\mha$ cannot be much larger than
$\mw$.\foot{If $\mha\sim {\cal O}(\msusy)$, then $H^\pm$, $\hh$ and
$\ha$ would all have masses of order $\msusy$, and the effective
low-energy theory below $\msusy$ would be that of the minimal
Standard Model.  Clearly, the above computation would not be
appropriate in this case.}
Of course, eq.~\llform\ is only a one-loop result.  This result is
improved by using the full RGE solution to $\lambda_4(\mw^2)$
$$\mhpm^2=\mha^2-\half\lambda_4(\mw^2)(v_1^2+v_2^2)\,.\eqn\chiggsrge$$

\REF\turski{J.F. Gunion and A. Turski, {\sl Phys. Rev.}
{\bf D39} (1989) 2701; {\bf D40} (1989) 2333.}
\REF\berz{A. Brignole, J. Ellis, G. Ridolfi and F. Zwirner,
{\sl Phys. Lett.} {\bf B271} (1991) 123; [E: {\bf B273} (1991) 550].}
\REF\brignole{A. Brignole, {\sl Phys. Lett.} {\bf B277} (1992) 313.}
Although the leading-log formula for $\mhpm$ [eq.~\llform] gives
a useful indication as to the size of the radiative corrections,
non-leading logarithmic contributions can also be important
in certain regions of parameter space.  A more complete set of
radiative corrections can be found in the literature%
\refmark{\pokorski,\diaz--\brignole}.
In the numerical results to be
exhibited below, important non-leading corrections to
the charged Higgs mass are also included (as described in ref.~[\diaz]).
However, it should be emphasized that the radiative corrections
to the charged Higgs mass are significant only for $\tanb<1$, a
region of MSSM parameter space not favored in supersymmetric
models.

The computation of the neutral CP-even Higgs masses follows a
similar procedure.  The results
are summarized below\refmark\llog.  From eq.~\massmhh, we see that we
only need results for $\lambda_1$,
$\lambda_2$ and $\widetilde\lambda_3\equiv\lambda_3+\lambda_4+\lambda_5$.
(Recall that in the leading-log analysis,
$\lambda_5=\lambda_6=\lambda_7=0$ at all energy scales.)
By iterating the corresponding RGEs as before, we end up with
$$\eqalign{
\lambda_1(\mz^2)&=~~\fourth[g^2+g'^2](\mz^2)
+{g^4\over384\pi^2\cw^4}\Bigg[ P_t\ln\left({\msusyy\over
m_t^2}\right)\crr
&\qquad+\bigg(12N_c{m_b^4\over\mz^4\cb^4}-6N_c{m_b^2\over\mzz\cb^2}
+P_f+P_g+P_{2H} \bigg)\ln\left({\msusyy\over
m_Z^2}\right)\Bigg]\,,\crr \crr
\lambda_2(\mz^2)&=~~\fourth
[g^2+g'^2](\mz^2)
+{g^4\over384\pi^2\cw^4}\Bigg[\bigg(P_f+P_g+P_{2H}
\bigg)\ln\left({\msusyy\over m_Z^2}\right) \crr
&\qquad+\bigg(12N_c{m_t^4\over\mz^4\sb^4}-6N_c{m_t^2\over\mzz\sb^2}
+P_t\bigg)\ln\left({\msusyy\over m_t^2}\right)\Bigg]\,,\crr\crr
\widetilde\lambda_3(\mz^2)&=-\fourth[g^2+g'^2](\mz^2)
-{g^4\over384\pi^2\cw^4}\Bigg[\bigg(-3N_c{m_t^2\over\mzz\sb^2}
+P_t\bigg)\ln\left({\msusyy\over m_t^2}\right)\crr
&\qquad+\bigg(-3N_c{m_b^2\over\mzz\cb^2}+P_f+P_g'+P_{2H}'
\bigg)\ln\left({\msusyy\over m_Z^2}\right)\Bigg]
\,,\cr}\eqn\dlambda$$
where
$$\eqalign{P_t~&\equiv~~N_c(1-4e_u\sw^2+8e_u^2\sw^4)\,,\cr
P_f~&\equiv~~ N_g\bigg\{N_c\left[2-4\sw^2+8(e_d^2+e_u^2)\sw^4\right]
+[2-4\sw^2+8\sw^4]\bigg\}-P_t\,,\cr
P_g~&\equiv-44+106\sww-62\sw^4\,,\cr
P_g'~&\equiv~~10+34\sww-26\sw^4\,,\cr P_{2H}&\equiv
-10+2\sww-2\sw^4\,,\cr
P_{2H}'&\equiv~~8-22\sww+10\sw^4\,.\cr}\eqn\defpp$$
In the above formulae, the electric charges of the quarks are $e_u
= 2/3$, $e_d = -1/3$, and the subscripts $t, f, g$ and $2H$
indicate that these are the contributions from the
top quark, the fermions (leptons and quarks excluding the top quark),
the gauge bosons and the two Higgs doublets (and corresponding
supersymmetric partners), respectively.
As in the derivation of $\lambda_4(\mw^2)$ above, we have improved
our analysis by removing the effects of top-quark loops below
$\mu=\mt$.  This requires a careful treatment of the evolution of
$g$ and $g'$ at scales below $\mu=\mt$.  The correct procedure is
somewhat subtle, since the full electroweak gauge symmetry is
broken below top-quark threshold; for further details, see
ref.~[\llog].  However, the following pedestrian
technique works: consider the RGE for $g^2+g'^2$ valid for $\mu<\msusy$
$${d\over dt}(g^2+g'^2)={1\over 96\pi^2}\Big[\left(8g^4
+\fortythirds g'^4\right)N_g+(g^4+g'^4)N_H-44g^4\Big]\,.\eqn\grge$$
This equation is used to run $g^2+g'^2$,
which appears in eq.~\boundary, from $\msusy$ down to $\mz$.
As before, we identify the term
proportional to $N_g$ as corresponding to the fermion loops.
We can explicitly extract the $t$-quark contribution by noting that
$$\eqalign{%
N_g\left(8g^4+\fortythirds g'^4\right)&=
{g^4N_g\over\cw^4}
\Big[\sixtyfourthirds s_W^4-16 s_W^2+8\Big]\crr
&={g^4\over\cw^4} \Big\{N_c\left[1+(N_g-1)\right] (1-4e_u s_W^2+8e_u^2
s_W^4)\crr
&\qquad + N_c N_g(1+4e_d s_W^2+8e_d^2 s_W^4)
+N_g(2-4 s_W^2+8s_W^4)\Big\}\,,\cr}
\eqn\pedest$$
\vskip12pt\noindent
where in the first line of the last expression, the term proportional
to 1 corresponds to the $t$-quark
contribution while the term proportional
to $N_g-1$ accounts for the $u$ and $c$-quarks; the second line contains
the contributions from the down-type quarks and leptons respectively.
Thus, iterating to one-loop,
$$\eqalign{%
(g^2+g'^2)(\msusy^2)&=
(g^2+g'^2)(\mz^2)+{g^4\over 96\pi^2 c_W^4}
\Bigg\{P_t\ln\left({\msusy^2\over\mt^2}\right)\crr
&\qquad+\left[P_f+(\sw^4+\cw^4)N_H-44\cw^4\right]
\ln\left({\msusy^2\over\mz^2}
\right)\Bigg\}\,.\cr}\eqn\gaugeiter$$
This result and terms that are proportional to $\mt^2$ and $\mt^4$
yield the terms in eq.~\dlambda\ that contain $\ln(\msusy^2/\mt^2)$.

The final step is to insert the expressions obtained in eq.~\dlambda\
into eq.~\massmhh.  The resulting
matrix elements for the mass-squared matrix to one-loop leading
logarithmic accuracy are given by
$$\eqalign{\calm_{11}^2&=\mha^2\sb^2+m_Z^2\cb^2
+{g^2\mzz\cb^2\over96\pi^2\cw^2}\Bigg[
P_t~\ln\left({\msusyy\over m_t^2}\right)\crrr
&\quad+\bigg(12N_c{m_b^4\over\mz^4\cb^4}-6N_c{m_b^2\over\mzz\cb^2}
+P_f+P_g+P_{2H} \bigg)\ln\left({\msusyy\over
m_Z^2}\right)\Bigg] \crr   \crr
\calm_{22}^2&=\mha^2\cb^2+m_Z^2\sb^2
+{g^2\mzz\sb^2\over96\pi^2\cw^2}\Bigg[\bigg(P_f+P_g+P_{2H}
\bigg)\ln\left({\msusyy\over m_Z^2}\right)\crrr
&\quad+\bigg(12N_c{m_t^4\over\mz^4\sb^4}-6N_c{m_t^2\over\mzz\sb^2}
+P_t\bigg)\ln\left({\msusyy\over m_t^2}\right)\Bigg]\crr \crr
\calm_{12}^2&=-\sb\cb\Biggl\{\mha^2+m_Z^2
+{g^2\mzz\over96\pi^2\cw^2}\Bigg[\bigg(P_t-3N_c{m_t^2\over\mzz\sb^2}
\bigg)\ln\left({\msusyy\over m_t^2}\right)\crrr
&\quad+\bigg(-3N_c{m_b^2\over\mzz\cb^2}+P_f+P_g'+P_{2H}'
\bigg)\ln\left({\msusyy\over
m_Z^2}\right)\Bigg]\Biggr\}\,.\cr}\eqn\mtophree$$
\vskip5pt\noindent
Diagonalizing this matrix [eq.~\mtophree] yields
the radiatively corrected CP-even Higgs masses and mixing angle $\alpha$.
%One can check that
%if $m_b=0$ and $\sin\beta=1$, then $\mhl^2={\cal M}_{22}^2$
%reproduces the leading logarithmic terms
%of eq.~\shiftapprox\ (after putting
%$M_{\widetilde Q}=M_{\widetilde\chi}=\msusy$ and $\mha=\mz$).

\REF\chengli{T.P. Cheng, E. Eichten and L.-F. Li, {\sl Phys. Rev.}
{\bf D9} (1974) 2259.}
\REF\cmpp{N. Cabibbo, L. Maiani, G. Parisi, R. Petronzio
\sl Nucl. Phys. \bf B158 \rm (1979) 295.}
The analysis presented above assumes that $\mha$ is not much larger
than ${\cal O}(\mz)$ so that the Higgs sector of the
low-energy effective theory contains the full two-Higgs-doublet spectrum.
On the
other hand, if $\mha\gg\mz$, then only $\hl$ remains in the low-energy
theory.  In this case, we must integrate
out the heavy Higgs doublet, in which case
one of the mass eigenvalues of ${\cal M}_{ij}^2$ is much
larger than the weak scale. In order to obtain the effective
Lagrangian at $\mweak$, we first have to run the various coupling
constants to the threshold $\mha$. Then we diagonalize the
Higgs mass matrix and express the Lagrangian in terms of the mass
eigenstates. Notice that in this case the mass eigenstate $\hl$ is
directly related to the field with the non-zero
vacuum expectation value [\ie,
$\beta(\mha)=\alpha(\mha)+\pi/2+{\cal O}(\mz^2/\mha^2)$]. Below $\mha$
there remains only the Standard Model Higgs doublet
$\phi\equiv\cb\Phi_1+\sb\Phi_2$. The potential is
$$
{\cal V}=m_{\phi}^2(\phi^{\dagger}\phi)
+\half\lambda(\phi^{\dagger}\phi)^2
\,,\eqn\potsm
$$
\vskip5pt\noindent
and the light CP-even Higgs mass is obtained using
$\mhl^2=\lambda v^2$.
The RGE in the Standard Model for $\lambda$ is\refmark{\chengli,\cmpp}\
\vskip5pt
$$
16\pi^2\beta_{\lambda} = 6\lambda^2
+\threeighth \left[2g^4+(g^2+g'^2)^2\right]-2\sum_i
N_{c_i}h_{f_i}^4 -\lambda\biggl(\ninehalf
g^2+\threehalf g'^2-2\sum_i N_{c_i}
h_{f_i}^2\biggr),
\eqn\defbetl $$
\vskip12pt\noindent
where the summation is over all fermions with
$h_{f_i}=gm_{f_i}\big/(\sqrt{2}\mw)$. The RGEs for the gauge couplings
are obtained from $\beta_{g^2}$ and $\beta_{g'^2}$
given in Appendix B by putting
$N_H=1$.  In addition, we require
the boundary condition for $\lambda$ at $\mha$
\vskip5pt
$$
\eqalign{\lambda(\mha)&=\left[\cb^4\lambda_1+\sb^4\lambda_2+
2\sb^2\cb^2(\lambda_3+\lambda_4+\lambda_5)
+4\cb^3\sb\lambda_6+4\cb\sb^3\lambda_7\right](\mha)\crrr
&=\left[\fourth(g^2+g'^2)\ctwob^2\right](\mha)
+{g^4\over384\pi^2\cw^4} \ln\left({\msusyy\over\mha^2}\right)\crrr
&~~\times\bigg[12N_c\bigg({m_t^4\over\mz^4}+{m_b^4\over\mz^4}\bigg)
+6N_c\ctwob\bigg({m_t^2\over\mzz}-{m_b^2\over\mzz}\bigg)\crrr
&\qquad+\ctwob^2\big(P_t+P_f\big)+(\sb^4+\cb^4)(P_g+P_{2H})
-2\sb^2\cb^2(P_g'+P_{2H}')\bigg]\,,
} \eqn\lapprox$$
\vskip12pt\noindent
where $(g^2+g'^2)c_{2\beta}^2$ is to be evaluated at the scale
$\mha$ as indicated.  The RGE for $g^2+g'^2$ was given in eq.~\grge;
note that at scales below $m_A$ we must set $N_H=1$.  Finally,
we must deal with implicit scale dependence of $c_{2\beta}^2$.
Since the fields $\Phi_i$ $(i=1,2)$ change with the scale, it
follows that $\tanb$ scales like the ratio of the two Higgs doublet
fields, \ie,
\vskip5pt
$$
{1\over\tan^2\beta}{d\tan^2\beta\over
dt}={\Phi_1^2\over\Phi_2^2}{d\over dt}
\left({\Phi_2^2\over\Phi_1^2}\right)
=\gamma_2-\gamma_1\,.\eqn\deftanb $$
\vskip5pt\noindent
Thus we arrive at the RGE for $\cos 2\beta$ in terms of the
anomalous dimensions $\gamma_i$ given in eq.~\wavez.
Solving this equation iteratively to first order yields
\vskip5pt
$$
\ctwob^2(\mha)=\ctwob^2(\mz)
+4\ctwob\cb^2\sb^2(\gamma_1-\gamma_2)
\ln\left({\mha^2\over\mzz}\right)\,.\eqn\rgecb $$
\vskip5pt

The one loop leading log expression for $\mhl^2 = \lambda(\mz)
v^2$ can now be obtained by solving the RGEs above for $\lambda(\mz)$
iteratively to first order using the boundary condition given
in eq.~\lapprox.  The result is
\vskip5pt
$$
\eqalign{
\mhl^2&= \mzz\ctwob^2(\mz)
+{g^2m_Z^2\over96\pi^2\cw^2}
\Bigg\{\bigg[12N_c{m_b^4\over\mz^4}-6N_c\ctwob{m_b^2\over\mzz}
+\ctwob^2P_f\crr
&~~+\left(P_{g}+P_{2H})(\sb^4+\cb^4\right)
 -2\sb^2\cb^2\left(P_{g}'+P_{2H}'\right)
 \bigg]\ln\left({\msusyy\over\mzz} \right)\crr
&~~+\bigg[12N_c{m_t^4\over\mz^4}+
 6N_c\ctwob{m_t^2\over\mzz}+\ctwob^2P_t\bigg]
 \ln\left({\msusyy\over m_t ^2}\right)\crr
&~~-\bigg[\left(\cb^4+\sb^4\right)P_{2H}-2\cb^2\sb^2P_{2H}'-P_{1H}\bigg]
 \ln\left({\mha^2\over\mzz}\right)\Bigg\}
 +{\cal O}\left({\mz^4\over\mha^2}\right)\,,
}\eqn\mhltot$$
\vskip5pt\noindent
where the term proportional to
\vskip5pt
$$
P_{1H} \equiv -9\ctwob^4+(1-2\sww+2\sw^4)\ctwob^2\,,
\eqn\defpps$$
\vskip5pt\noindent
corresponds to the Higgs
boson contribution in the one-Higgs-doublet model.
%The term in eq.~\mhltot\ proportional to $\ln(\mha^2)$
%accounts for the fact that there are two Higgs doublets present at
%a scale above $\mha$ but only one Higgs doublet below $\mha$.
%As a check, one can verify that eq.~\shiftapprox\ is reproduced in
%the limit of large $\tanb$ (and $m_b=0$).

The leading-log
formulae presented above are expected to be accurate as long
as: (i) there is one
scale characterizing supersymmetric masses, $\msusy$, which is
large and sufficiently separated from $\mz$
(say, $\msusy\gsim 500$~GeV),
(ii) $m_t$ is somewhat above $\mz$ (say, $\mt\gsim 125$~GeV)
while still being small compared to $\msusy$, and (iii)
the squark mixing parameters are not unduly large.  In particular,
(ii) is an important condition---it is the dominance of the leading
$\mt^4\ln(\msusy^2/\mt^2)$ term that guarantees that the
non-leading logarithmic terms are unimportant.
\foot{In contrast, there
is no leading logarithmic contribution to $\mhpm^2$ that grows with
$\mt^4$.  As a result, the non-leading logarithmic terms tend to be
more important as discussed earlier.}
Under these conditions, the largest non-leading logarithmic term is
of ${\cal O}(g^2\mt^2)$, which can be identified from a full one-loop
computation as being the subdominant term relative to the leading
${\cal O}(g^2\mt^4\ln\msusy^2)$ term in ${\cal M}_{22}^2$.  Thus,
we can make a minor improvement on our computation of the leading-log
CP-even Higgs squared mass matrix by taking
\vskip5pt
$$
{\cal M}^2 = {\cal M}^2_{\rm LL} +
{N_c g^2\mt^2\over48\pi^2\sb^2c_W^2}\left(\matrix{0&0\cr0&1}\right)
\eqn\nllogtrm
$$
\vskip10pt\noindent
where $\calm^2_{\rm LL}$ is the leading-log CP-even Higgs squared mass
matrix [given to one-loop in eq.~\mtophree].
%For $\beta=\pi/2$, this
%correction yields the non-leading-logarithmic term  displayed in
%eq.~\shiftapprox.
The shift of the
light Higgs mass due to this non-leading-log correction is of
order 1 GeV.  Finally, the case of multiple and widely separated
supersymmetric particle thresholds and/or
large squark mixing (which is most likely in the top squark sector),
leads to
new non-leading logarithmic contributions to the scalar mass-squared
matrix can become important.  These are discussed in refs.~[\llog] and
[\hemperice].

\FIG\mhtbvma{%
RGE-improved Higgs mass $\mhl$ as a function
of $\tanb$ for (a) $\mt = 150$ GeV and (b) $\mt = 200$ GeV.
Various
curves correspond to $\mha = 0,~20,~50,~100$ and $300$ GeV as labeled in
the figure. All $A$-parameters and $\mu$ are set equal to zero.
The light CP-even Higgs mass varies very weakly with $\mha$
for $\mha>300$ GeV.  Taken from ref.~[\llog].}
\bigskip\bigskip
\leftline{{\bf 6. Implications of the Radiatively Corrected MSSM Higgs
Sector}}
\medskip

In this section I shall briefly survey some of the numerical
results for the radiatively corrected Higgs masses and couplings.
Additional results can be found in ref.~[\llog].
Complementary work can be found in refs.~[%
\radmssm,%{25}
\moreradmssm,% {26}
\pokorski,%{28}
\berkeley--%{30}
\brignole].  %{37}
In fig.~\mhtbvma (a) and (b) I
plot the light CP-even Higgs mass as a function of
$\tanb$ for $\mt = 150$ and 200 GeV for various choices of $\mha$.
All $A$-parameters and $\mu$ are set equal to zero.
Perhaps the most dramatic consequence of the radiatively corrected
Higgs sector is the large violation of the tree-level bound
$\mhl\leq\mz$.  The new $\mhl$ bound ($\mhl^{\rm max}$)
is saturated when $\mha$ and $\tanb$ are large.  Moreover, when
$\mha\geq\mhl^{\rm max}$, one sees that $\mhl$ is large throughout
the entire $\tanb$ region;
the radiatively corrected $\mhl$ reaches
a maximum (minimum) at $\tanb\simeq\infty$ $(\tanb\simeq1)$.
In particular, there is a substantial
region of Higgs parameter space in which $\mhl$ lies above the
current LEP experimental Higgs mass limits.
Indeed, for $\msusy=1$~TeV, $\mt=200$~GeV,
and $\mha\gsim 200$~GeV, fig.~\mhtbvma(b) indicates that
$\mhl>\mz$ independent of the
value of $\tanb$.  Thus, there is a non-negligible region of parameter
space in which the $\hl$ is kinematically inaccessible to LEP-II
(running at $\sqrt{s}\leq 200$~GeV).  This is quite a departure from
the tree-level expectations of $\mhl\leq\mz$ in which the LEP
discovery of the Higgs boson was assured if the MSSM were correct.

%\midinsert
%   \tenpoint \baselineskip=12pt   \narrower
%\plotpicture{\hsize}{9cm}{hmass.topdraw}
%\vskip6pt\noindent
%{\bf Fig.~\mhtbvma.}\enskip
%\endinsert

Nevertheless, the LEP Higgs search does rule out some
regions of the Higgs parameter space.  These regions typically
correspond to smaller values of $\mha$ where $\hl$ can be
relatively light.
For fixed $\tanb$, $\mhl$ reaches its minimum
value, $\mhl^{\rm min}$, when $\mha\to 0$.  In contrast to the
tree-level behavior (where $\mhl\leq \mha$),
the Higgs mass does not vanish
as $\mha\to 0$. Moreover, $\mhl^{\rm min}$
increases as $\tanb$ decreases but
exhibits only a moderate dependence on $\mt$ and $\msusy$.
One interesting consequence is that
there exists a range of parameters for which
the tree-level bound, $\mhl\leq\mha$ is violated.  In fact,
the results of fig.~\mhtbvma\ indicate that in the region of small
$\tanb$ and small $\mha$, it is possible to have $\mhl>2\mha$, thereby
allowing a new decay-mode $\hl\to\ha\ha$ which is kinematically
forbidden at tree-level.

\REF\colemanweinberg{S. Coleman and E. Weinberg, {\sl Phys.~Rev.}
{\bf D7} (1973) 1888.}
\REF\hhgreftwo{See section 2.5 of ref.~[\hhg].}
\REF\franco{E. Franco and A. Morelli, {\sl Nuovo Cim.} {\bf 96A} (1986)
257.}
One other difference between the tree-level prediction for $\mhl$ and the
results of fig.~\mhtbvma\ is noteworthy.   From eq.~\kviii, we see
that for $\tanb=1$, $\mhl=0$ at tree-level.  The results of
fig.~\mhtbvma\ indicate that the radiative corrections to $\mhl$ are
substantial for $\tanb=1$, particularly when $\mt$ is large.  This
is again a consequence of the $g^2 m_t^4\ln(\msusy^2/\mz^2)$ enhancement
of ${\cal M}_{22}^2$.
The $\tanb=1$ limit is analogous to the
Coleman-Weinberg limit\refmark\colemanweinberg\ of
the Standard Model, in which the mass of the Higgs boson arises
entirely from radiative corrections.  However, in the Standard Model,
the Coleman-Weinberg mechanism cannot be operative
if $\mt\gsim\mw$\refmark\hhgreftwo\ (and
in any case, the Higgs mass that arises from this mechanism cannot be
larger than about 10 GeV, which is ruled
out by the LEP Higgs search\refmark\lepsearch).
Clearly, no such restriction exists in the
MSSM\refmark{\franco,\diaztwo}.  The difference lies
in the large positive contribution to the Higgs squared mass from
a loop of top squarks.  From fig.~\mhtbvma(b), we see that for
$\mt=200$~GeV and $\tanb=1$, a value of $\mhl$ as large as 100~GeV is
possible.  Thus, LEP cannot yet rule out the possibility that the mass
of the lightest CP-even Higgs boson arises entirely from radiative
corrections\refmark\diaztwo.

%\topinsert
%   \tenpoint \baselineskip=12pt   \narrower
%\plotpicture{\hsize}{9cm}{tregim.topdraw}
%\vskip6pt\noindent
%{\bf Fig.~\regime.}\enskip
\FIG\regime{%
The range of allowed Higgs masses for
large $\mha$ (in these plots, $\mha=300$ GeV).
The lower limit corresponds to $\tanb=1$.
The upper limit corresponds to the limit of large
$\tanb$ (we take $\tanb=20$). In (a) and (b)
$\mt$ is varied for $\msusy= 1$ and $0.5$
TeV, respectively. In (c) and (d)
$\msusy$ is varied and $\mt= 150$ and $200$~GeV,
respectively. The solid (dashed) curves in (c) and (d)
correspond to the computation in which the RGEs are solved numerically
(iteratively to one-loop order).  Taken from ref.~[\llog].}
%\vskip15pt
%\endinsert

In the limit $\mha\to\infty$, the couplings of $\hl$ to gauge
bosons and matter fields are identical to the Higgs couplings of
the Standard Model so that the Higgs sector of the two models cannot be
phenomenologically distinguished.
However, supersymmetry does impose constraints on the quartic Higgs
self-coupling at the scale $\msusy$, and this  influences
the possible values of $\mhl$. To illustrate this point, I have plotted
in fig.~\regime\
the range of allowed $\mhl$ in the case of large $\mha$ (taken here to
be $\mha=300$ GeV).
As noted above, the lower limit for $\mhl$ is attained
if $\tanb\simeq1$ and the upper limit is attained in the limit of large
$\tanb$ (taken to be $\tanb=20$ in fig.~\regime).\foot{A second
maximum for $\mhl$ would arise for very
small $\tanb$; however, this lies outside the permitted region indicated
in figs.~\tanlima\ and \tanlimb.}
Suppose the
top quark mass is known
and that $\hl$ is discovered with Standard Model couplings. If
$\mhl$ does not lie in the allowed mass region displayed in fig.~\regime,
one could conclude that the MSSM is ruled out.
Fig.~\regime\ also exhibits the sensitivity to the choice of $\msusy$.
The larger the value of $\msusy$, the more significant the corrections
to the Higgs mass due to full renormalization group improvement.
In fig.~\regime(c) and (d), the dashed lines (labeled 1LL for one-loop
leading-log) correspond to computing $\mhl$ by exactly diagonalizing the
squared mass matrix given in eq.~\mtophree.  The solid lines (denoted
by RGE) are obtained by solving numerically the RGEs for the $\lambda_i$,
inserting the results into eq.~\massmhh, and computing the eigenvalue of
the lighter CP-even Higgs scalar.  For $\msusy=1$~TeV, the largest
discrepancy between the RGE and 1LL results occurs for
large $\mt$ and $\mha$.  For example, for $\tanb=1$, $\mha=300$~GeV
and $\mt = 200$~GeV, we find $(\mhl)_{\rm RGE} = 96.8$~GeV
while $(\mhl)_{\rm 1LL} = 104.4$~GeV.   Values of $\msusy$ much larger
than 1~TeV would be in conflict with the philosophy of low-energy
supersymmetry.

%\topinsert
%   \tenpoint \baselineskip=12pt   \narrower
%\plotpicture{\hsize}{9cm}{hmassrc1.topdraw}
%\vskip6pt\noindent
%{\bf Fig.~\massesra.}\enskip
\FIG\massesra{%
The masses of $\hl$, $\hh$ and $\hpm$ in the MSSM
for $\mha=50$ and 200~GeV.
The neutral CP-even Higgs masses are obtained from a calculation
that includes the leading-log one-loop radiative corrections
[based on eq.~\mtophree].
The charged Higgs mass is obtained from a similar calculation, but
important non-leading logarithmic effects have also been included%
\refmark\diaz.
All supersymmetric masses are assumed to be roughly degenerate
of order $\msusy=1$~TeV.
The two curves for each Higgs mass shown
correspond to $\mt=150$ and 200 GeV. The larger neutral Higgs mass
corresponds to the larger $\mt$ choice.  In the case of $\hpm$,
$\mhpm$ increases [decreases] with $\mt$ for large [small] $\tanb$.}
%
%\endinsert

Consider next the predictions of the one-loop radiatively
corrected Higgs sector for the other physical Higgs bosons of the MSSM.
In fig.~\massesra, I plot the radiatively corrected MSSM Higgs masses
as a function of $\tanb$ for $\msusy=1$~TeV and for
two choices of $\mt$ and $\mha$.  (As above, all $A$ and $\mu$ parameters
are set to zero.)  The neutral Higgs masses have been obtained by
diagonalizing eq.~\mtophree.  Full RGE-improvement, which is not
included in fig.~\massesra, would change these results by no more than
about $5\%$.  In the case of the charged Higgs mass, important
non-leading logarithmic contributions have also been included, as
described in ref.~[\diaz].  Note that
the tree-level bound $\mhpm\geq\mw$ can
be violated, but only if $\tanb\lsim 0.5$ and $\mha$ is small.
The small $\tanb$ region corresponds to an enhanced
Higgs-top quark Yukawa coupling.  This also explains the increase of
$\mhh$ in this region, which is being controlled by the $\mt^4/\sb^2$
factor in ${\cal M}_{22}^2$ [eq.~\mtophree].
Of course, this same
factor is responsible for the violation of the bound $\mhl\leq\mz$.

\REF\damien{D. Pierce and A. Papadopoulos, {\sl
Phys. Rev.} {\bf D47} (1992) 222.}
Let us now turn to the impact of the radiative corrections on the
Higgs couplings.
To obtain radiatively corrected couplings that are accurate in the
leading logarithmic approximation, it is sufficient to use the
tree-level couplings in which the parameters are taken to be running
parameters evaluated at the electroweak scale.\foot{Once again, the
reader should be cautioned that non-leading-logarithmic
radiative corrections may be significant in certain
parameter regimes.  See ref.~[\damien] for an illustration of this
point in the case of the radiatively corrected $\hl ZZ$ vertex in the
MSSM.}
First, recall that
$\tanb$ and $\mha$ are input parameters.  Next, we obtain
the CP-even Higgs mixing angle $\alpha$ by diagonalizing the
radiatively corrected CP-even Higgs mass matrix.
With the angle $\alpha$ in hand one may compute, for example,
$\cos(\beta-\alpha)$ and $\sin\alpha$.  These results can be used
to obtain the Higgs couplings
to gauge bosons [eq.~\littletable] and fermions [eq.~\qqcouplings].
Finally, the Higgs self-couplings [eq.~\defghaa] are obtained by
making use of the $\lambda_i$ evaluated at the electroweak scale.
The end result is a complete
set of Higgs boson decay widths and branching ratios that include
leading-log radiative corrections.

\FIG\cbafunction{%
The factor $\cos^2(\beta-\alpha)$ as a function of
$\mha$ for $\tanb = 0.5, 1 ,2$ and 20 (dotted, dashed, dot-dashed and
solid curves, respectively). Results are presented for $\msusy=1$ TeV. We
consider the case of (a) $\mt=150$ GeV and (b) $\mt=200$ GeV.
Taken from ref.~[\llog].}

The Higgs production cross-section
in a two-Higgs-doublet model via the process
$e^+e^-\to Z\to Z\hh(Z\hl)$ is suppressed by a
factor $\cos^2(\beta-\alpha)$ [$\sin^2(\beta-\alpha)$]
as compared to the corresponding cross-sections in the
Standard Model.
In fig.~\cbafunction\ I plot $\cos^2(\beta-\alpha)$ as a
function of $\mha$ for $\tanb = 0.5, 1, 2$ and 20,
for $A_t=A_b=\mu=0$,
$\msusy=1$ TeV and two choices of $\mt$.
The behavior is similar to that of the tree-level result:
(i) for $\mha\ll\mz$ and
$\tan\beta\gg 1$, $\cos^2(\beta-\alpha)\simeq 1$, and
(ii) for $\mha\gsim 2\mz$,
$\sin^2(\beta-\alpha)\simeq 1$. The fact that
$\cos^2(\beta-\alpha)\to 0$ as $\mha$ becomes large
is expected
since for large $\mha$, all heavy Higgs states decouple, while the
$\hl ZZ$ coupling [which is proportional to $\sin(\beta-\alpha)$]
approaches its Standard Model value.
Nevertheless, it is interesting to note that
$\cos^2(\beta-\alpha)$ approaches 0 more slowly as $\mt$ increases
(\ie, as the radiative corrections become more significant).

%\topinsert
%   \tenpoint \baselineskip=12pt   \narrower
%\plotpicture{\hsize}{9cm}{cba.topdraw}
%\vskip6pt\noindent
%{\bf Fig.~\cbafunction.}\enskip
%
%\endinsert

\REF\nirtwo{H.E. Haber, R. Hempfling and Y. Nir,
{\sl Phys. Rev.} {\bf D46} (1992) 3015.}
%\topinsert
%   \tenpoint \baselineskip=12pt   \narrower
%\plotpicture{\hsize}{9cm}{haabr3.topdraw}
%\vskip6pt\noindent
%{\bf Fig.~\haabrd.}\enskip
\FIG\haabrd{%
Regions of nonvanishing $BR(\hl\to\ha\ha)$
%in the $\tan\beta$---$m_t$ plane
for $\mha=5, 10, 20$ and 30~GeV. To the right of
the solid curves, $\mhl<2\mha$, and the decay $\hl\to\ha\ha$
is kinematically forbidden.  To the left of
the dashed curve, $BR(\hl\to \ha\ha)\geq 0.5$ and
between the dotted curves, $BR(\hl\to \ha\ha)\geq
0.8$.  $\msusy=1$~TeV in all four graphs.
Taken from ref.~[\nirtwo].}
%\vskip15pt
%\endinsert

When radiative corrections have been incorporated, new possibilities
arise that did not exist at tree-level.  One example mentioned
earlier is the possibility of the decay $\hl\to\ha\ha$,
which is kinematically forbidden at tree-level but allowed for some
range of MSSM parameters\refmark{\berz,\nirtwo}.
We can obtain the complete one-loop leading-log
expression for the $\hl\ha\ha$ coupling (assuming $\mha \lsim m_Z$)
by inserting
the one-loop leading-log formulae for the $\lambda_i$ into
eq.~\defghaa\refmark\nirtwo\
$$\eqalign{%
&{g_{\hl\ha\ha}\over
g m_Z/2\cw}=-c_{2\beta}s_{\beta+\alpha}\left\{
1+{g^2\over96\pi^2\cw^2}\left[P_t\ln \!
\left({\msusyy\over m_t^2}\right)+P_f\ln\!\left({\msusyy\over\mzz}\right)
\right]\right\}\crr\crr
&~~+{g^2N_c\over16\pi^2m_W^2m_Z^2}\left\{\left[
{\sa\sb^2\over\cb^3}(2m_b^4-m_b^2m_Z^2\cb^2)
-{(\ca\sb^3-\sa\cb^3)\over2\cb^2}m_b^2m_Z^2\right]
\ln\!\left({\msusyy\over m_Z^2}\right)\right.\crr\crr
&~~-\left.\left[{\ca\cb^2\over\sb^3}(2m_t^4-m_t^2m_Z^2\sb^2)
+{(\ca\sb^3-\sa\cb^3)\over2\sb^2}m_t^2m_Z^2\right]
\ln\!\left({\msusyy\over m_t^2}\right)\right\}\crr\crr
&~~-{g^2\over192\pi^2\cw^2}\left[s_{2\beta}c_{\beta+\alpha}
(P_{2H}+P_g)
-2(\ca\sb^3-\sa\cb^3)(P_{2H}'+P_g')\right]\ln\!\left({\msusyy\over
m_Z^2}\right)\,.\cr}\eqn\ghaall$$
\vskip5pt\noindent
Once kinematically allowed, $\hl\to\ha\ha$ is almost
certainly the dominant decay mode as shown in fig.~\haabrd.
%These results indicate the importance of the
%search for $\hl\to\ha\ha$ at LEP.
As $\mha$ increases beyond 30 GeV,
the region of parameter space  where this decay is
permitted quickly shrinks.

\REF\gbhs{J.F. Gunion, R. Bork, H.E. Haber and A. Seiden,
{\sl Phys. Rev.} {\bf D46} (1992) 2040;
J.F. Gunion, H.E. Haber and C. Kao,
{\sl Phys. Rev.} {\bf D46} (1992) 2907;
V. Barger, M.S. Berger, A.L. Stange, and R.J.N. Phillips,
{\sl Phys. Rev.} {\bf D45} (1991) 4128;
Z. Kunszt and F. Zwirner, {\sl Nucl. Phys.} {\bf B385} (1992) 3;
A. Yamada, {\sl Mod. Phys. Lett.} {\bf A7} (1992) 2877.}
\REF\brigz{A. Brignole and F. Zwirner, {\sl Phys. Lett.} {\bf B299}
(1993) 72.} %CERN-TH-6603-92 (1992).}
\REF\jack{J.F. Gunion, contribution to These Proceedings.}
\REF\finland{H.E. Haber, in {\it Physics and Experiments with Linear
Colliders},
Proceedings of the Workshop on Future $e^+e^-$ Linear Colliders,
Saariselk\"a, Finland, 9--14 September 1991, edited by R. Orava,
P. Eerola, and M. Nordberg (World Scientific, Singapore, 1992) p.~235.}
For the heavier Higgs states, there are many possible
final state decay modes.  The various branching ratios are complicated
functions of the MSSM parameter space\refmark{\gbhs,\brigz}.
Plots of the branching ratios of the MSSM Higgs bosons, with all one-loop
leading log corrections included, can be found in ref.~[\jack].  These
plots indicate a rich phenomenology for Higgs searches at
future colliders\refmark{\gbhs,\finland}.  Although the
possibility of a Higgs discovery at LEP still remains, the effects of
the radiative corrections (particularly if $\mt$ is near the upper end of
its expected range) suggest that the success of the Higgs Hunt must
await the supercollider era.
%Presumably, the SSC and LHC will uncover
%direct evidence for supersymmetric particles, if low-energy
%supersymmetry exists.  In this case, the details of the Higgs sector
%will contain crucial information regarding the structure of the
%theory---the mechanism of electroweak symmetry breaking and the nature
%of the TeV scale physics that lies beyond the Standard Model.

\bigskip
\leftline{{\bf 7. Final Remarks}}
\medskip

In this paper, I have attempted to survey the implications of
radiative corrections in the Minimal Supersymmetric Model (MSSM).
In precision electroweak measurements at LEP, one can examine both
oblique radiative corrections (where virtual heavy particle
effects enter only through gauge boson self-energies) and
non-oblique radiative corrections.
The dominant effects of new heavy particles to LEP observables
enter via the
oblique corrections and are neatly summarized by three numbers,
$S$, $T$ and $U$.  I have shown that the supersymmetric particle
contributions to these parameters are very small and vanish at least
as fast as $1/\msusy^2$ in the limit where the relevant MSSM particle
masses [assumed to be ${\cal O}(\msusy)$]
become sufficiently larger than $\mz$.
The results presented here imply that if low-energy
supersymmetry is realized in nature, then supersymmetric particles
will be directly discovered in future experiments before any deviation
from the Standard Model predictions for precision electroweak
measurements is detected at LEP.

Perhaps a fertile ground for MSSM-particle-mediated radiative
corrections lies in processes involving external $b$ and/or $t$ quarks.
In such processes, some non-oblique radiative corrections are
surprisingly large and may be experimentally observable.
An explicit example
discussed in this paper is the contribution of virtual charged Higgs
exchange to the rare radiative decay, $b\to s\gamma$.  If
present experimental results are improved by a factor of 2, very
powerful constraints on the properties of the charged Higgs mass
and other MSSM parameters will emerge.  For future research,
it will be interesting to consider similar effects in processes involving
external top-quarks.  In this regard, it is amusing to note that the
Eloisatron would be the ultimate top-quark factory, and would
perhaps permit the study of a variety of rare top-quark decays.

Finally, I showed that the largest (and perhaps the most dramatic)
set of MSSM radiative corrections can be found in the corrections
to the natural tree-level Higgs mass relations of the model.  These
corrections have led to a re-evaluation of MSSM Higgs phenomenology
at LEP and at future colliders.  In particular, LEP-200 is no longer
guaranteed to find the lightest CP-even Higgs boson even if the MSSM is
the correct low-energy theory.  Of course, these conclusions depend
sensitively on the value of the top quark mass.  Were CDF to announce
a top quark mass of, say, 120 GeV or less, then the impact of the
radiative corrections to the MSSM Higgs sector would be minimal.
On the other hand, with a heavier top quark becoming more likely
day by day, the significance of the radiatively corrected Higgs sector
becomes more apparent.

Radiative corrections, by their nature, are usually small corrections
to tree-level results.  Nevertheless, the detection of the radiative
corrections is crucial for the experimental verification of the
Standard Model as well as any successor.
If low-energy supersymmetry is correct, it will be crucial to test
the MSSM as precisely as possible.  The radiative corrections will
clearly play an important role in the verification of particle
theories for many years to come.
\endpage
\centerline{{\bf Acknowledgments}}
\medskip
I am grateful to Franco Anselmo, Luisa Cifarelli, Valery Khoze
and Antonio Zichichi for their invitation to a most stimulating
meeting, and to their hospitality and support during my stay in
Erice.  I also acknowledge my various collaborators---Riccardo
Barbieri, Marco Diaz, Ralf Hempfling and Yosef Nir, whose
contributions were instrumental to the development of much of the
material presented here.  In addition, I thank
Fabio Zwirner for his contributions to the work that produced
figs.~\tanlima\ and \tanlimb.  Finally, conversations with Dallas
Kennedy and Michael Peskin concerning $S$, $T$ and $U$ are gratefully
acknowledged.
This work was supported in part by the U.S. Department of Energy.

\vskip1.5in
\bigskip
\APPENDIX{A}{A:~~The Non-Supersymmetric Two-Higgs Doublet Model}
\medskip
Consider a model of
low-energy supersymmetry, where
the effective scale of supersymmetry breaking is
substantially larger than $\mz$, but where the
Higgs sector masses are of ${\cal O}(\mz)$.  In this
case, the effective low energy theory at energies below
the supersymmetry-breaking scale
corresponds to a general (non-supersymmetric) two-Higgs-doublet
model.  In principle, this model would contain CP-violating as well
as CP-conserving couplings.  For simplicity (and due to lack of knowledge
of the fundamental CP-violating parameters of the underlying
supersymmetric model), I will assume that all CP-violating effects
generated by integrating out the heavy supersymmetric particles are
small and can be neglected.

Thus, in this Appendix, I briefly summarize the key ingredients of
the most general
two-Higgs doublet extension of the Standard Model, assuming that
the Higgs sector conserves CP\refmark\hhgref.
Let $\Phi_1$ and
$\Phi_2$ denote two complex $Y=1$, SU(2)$\ls{L}$ doublet scalar fields.
The most general gauge invariant scalar potential is given by
\vskip5pt
$$\eqalign{
\calv&=m_{11}^2\Phi_1^\dagger\Phi_1+m_{22}^2\Phi_2^\dagger\Phi_2
-[m_{12}^2\Phi_1^\dagger\Phi_2+{\rm h.c.}]\crrr
&\quad +\half\lambda_1(\Phi_1^\dagger\Phi_1)^2
+\half\lambda_2(\Phi_2^\dagger\Phi_2)^2
+\lambda_3(\Phi_1^\dagger\Phi_1)(\Phi_2^\dagger\Phi_2)
+\lambda_4(\Phi_1^\dagger\Phi_2)(\Phi_2^\dagger\Phi_1)\crrr
&\quad +\left\{\half\lambda_5(\Phi_1^\dagger\Phi_2)^2
+\big[\lambda_6(\Phi_1^\dagger\Phi_1)
+\lambda_7(\Phi_2^\dagger\Phi_2)\big]
\Phi_1^\dagger\Phi_2+{\rm h.c.}\right\}\,.\cr}\eqn\pot$$
\vskip5pt\noindent
In most discussions of two-Higgs-doublet models, the terms proportional
to $\lambda_6$ and $\lambda_7$ are absent.  This can be achieved by
imposing a discrete symmetry $\Phi_1\to -\Phi_1$ on the model.  Such a
symmetry would also require $m_{12}=0$ unless we allow a
soft violation of this discrete symmetry by dimension-two terms.\foot{%
This latter requirement is sufficient to guarantee the absence of
Higgs-mediated tree-level flavor changing neutral currents.}
For the moment, I will refrain from setting any of the coefficients
in eq.~\pot\ to zero.  In principle, $m_{12}^2$, $\lambda_5$,
$\lambda_6$ and $\lambda_7$ can be complex.  However, I shall
ignore the possibility of CP-violating effects in the Higgs sector
by choosing all coefficients in eq.~\pot\ to be real.
The scalar fields will
develop non-zero vacuum expectation values if the mass matrix
$m_{ij}^2$ has at least one negative eigenvalue. Imposing CP invariance
and U(1)$\ls{\rm EM}$ gauge symmetry, the minimum of the potential is
$$\langle \Phi_1 \rangle={1\over\sqrt{2}}
\pmatrix{0\cr v_1\cr}, \qquad \langle \Phi_2\rangle=
{1\over\sqrt{2}}\pmatrix{0\cr v_2\cr}\,,\eqn\potmin$$
where the $v_i$ can be chosen to be real.  It is convenient to introduce
the following notation:
$$v^2\equiv v_1^2+v_2^2={4\mw^2\over g^2}\,,
\qquad\qquad\tb\equiv\tanb\equiv{v_2\over v_1}\,.\eqn\tanbdef$$
Of the original eight scalar degrees of freedom, three Goldstone
bosons ($G^\pm$ and $G^0$)
are absorbed (``eaten'') by the $W^\pm$ and $Z$.  The remaining
five physical Higgs particles are: two CP-even scalars ($\hl$ and
$\hh$, with $\mhl\leq \mhh$), one CP-odd scalar ($\ha$) and a charged
Higgs pair ($\hpm$). The mass parameters $m_{11}$ and $m_{22}$ can be
eliminated by minimizing the scalar potential.  The resulting
squared masses for the CP-odd and charged Higgs states are
\vskip4pt
$$\eqalign{%
\mha^2 &={m_{12}^2\over \sb\cb}-\half
v^2\big(2\lambda_5+\lambda_6\tb^{-1}+\lambda_7\tb\big)\,,\crrr
m_{H^{\pm}}^2 &=m_{A^0}^2+\half v^2(\lambda_5-\lambda_4)\,.\cr}
\eqn\mamthree$$
\vskip5pt\noindent
The two CP-even Higgs states mix according to the following squared mass
matrix:
$$\eqalign{%
\calm^2 &=m_{A^0}^2 \left(\matrix{\sb^2&-\sb\cb\cr
-\sb\cb&\cb^2}\right)\crrr
&+v^2\left( \matrix{\lambda_1\cb^2+2\lambda_6\sb\cb+\lambda_5\sb^2
 &(\lambda_3+\lambda_4)\sb\cb+\lambda_6
\cb^2+\lambda_7\sb^2\crr
(\lambda_3+\lambda_4)\sb\cb+\lambda_6
\cb^2+\lambda_7\sb^2&
\lambda_2\sb^2+2\lambda_7\sb\cb+\lambda_5\cb^2}\right)\,,\cr
}\eqn\massmhh$$
\vskip3pt\noindent
where $s_\beta\equiv\sin\beta$ and $c_\beta\equiv\cos\beta$.
The physical mass eigenstates are
$$\eqalign{%
\hh &=(\sqrt{2}{\rm Re\,}\Phi_1^0-v_1)\cos\alpha+
(\sqrt{2}{\rm Re\,}\Phi_2^0-v_2)\sin\alpha\,,\crr
\hl &=-(\sqrt{2}{\rm Re\,}\Phi_1^0-v_1)\sin\alpha+
(\sqrt{2}{\rm Re\,}\Phi_2^0-v_2)\cos\alpha\,.\cr}
\eqn\scalareigenstates$$
The corresponding masses are
$$
 m^2_{\hh,\hl}=\half\left[{\cal M}_{11}^2+{\cal M}_{22}^2
\pm \sqrt{({\cal M}_{11}^2-{\cal M}_{22}^2)^2 +4({\cal M}_{12}^2)^2}
\ \right]\,,
\eqn\higgsmasses$$
and the mixing angle $\alpha$ is obtained from
$$\eqalign{%
\sin 2\alpha &={2{\cal M}_{12}^2\over
\sqrt{({\cal M}_{11}^2-{\cal M}_{22}^2)^2 +4({\cal M}_{12}^2)^2}}\ ,\crrr
\cos 2\alpha &={{\cal M}_{11}^2-{\cal M}_{22}^2\over
\sqrt{({\cal M}_{11}^2-{\cal M}_{22}^2)^2 +4({\cal M}_{12}^2)^2}}\ .\cr}
\eqn\alphadef$$

\REF\hehnir{%
H.E. Haber and Y. Nir, {\sl Nucl. Phys.} {\bf B335} (1990) 363.}
\REF\ghw{%
J.M. Cornwall, D.N. Levin and G. Tiktopoulos, \PRL 30&73&1268&;
\PRB D10&74&1145&;
C.H. Llewellyn Smith, \PL 46B&73&233&;
H.A. Weldon, \PRB D30&84&1547&;
J.F. Gunion, H.E. Haber and J. Wudka, {\sl Phys. Rev.}
{\bf D43} (1991) 904.}

The phenomenology of the two-Higgs doublet model depends in detail on
the various couplings of the Higgs bosons to gauge bosons, Higgs bosons
and fermions.  The Higgs couplings to gauge bosons follow from
gauge invariance and are thus model independent.  For example, using
the symbol $\hsm$ for the minimal Higgs boson of the Standard Model,
the couplings of the two CP-even Higgs bosons to $W$ and $Z$ pairs
are given in terms of the angles $\alpha$ and $\beta$ by
$$\eqalign{g\ls{\hl VV}&=g\ls{\hsm VV}\sinbma \crr
           g\ls{\hh VV}&=g\ls{\hsm VV}\cosbma\,,\cr}\eqn\vvcoup$$
where $g\ls{\hsm VV}\equiv g \ls V m \ls V$ and
$$g\ls V=\cases{g,& $V=W\,$,\cr g/\cos\theta_W,& $V=Z\,$.}\eqn\hix
$$
There are no tree-level couplings of $\ha$ or $\hpm$ to $VV$.
Gauge invariance also determines the strength of the trilinear
couplings of one gauge boson to two Higgs bosons.  For example,
$$\eqalign{g\ls{\hl\ha Z}&={g\cosbma\over 2\cos\theta_W}\,,\crrr
           g\ls{\hh\ha Z}&={-g\sinbma\over 2\cos\theta_W}\,.
           \cr}\eqn\hvcoup$$

In the examples shown above, some of the couplings can be suppressed
if either $\sin(\beta-\alpha)$ or $\cos(\beta-\alpha)$ is very small.
If the theory yields a Higgs mass spectrum in which the $\hl$ is
considerably lighter than the other physical Higgs scalars, then
the $\hl$ couplings will typically approach their Standard Model
values\refmark\hehnir.
In this limit, $\cos(\beta-\alpha)\to 0$.
More generally, all the Higgs couplings
cannot vanish simultaneously. From the expressions above, we see
that the following sum rules must hold separately for $V=W$ and $Z$:
$$\eqalign{%
g_{\hh V V}^2 + g_{\hl V V}^2 &=
                      g_{\phi^0 V V}^2\,,\crr
g_{\hl\ha Z}^2+g_{\hh\ha Z}^2&=
             {g^2\over 4\cos^2\theta_W}\,.\cr}
\eqn\sumruletwo$$
These results are a
consequence of the tree-unitarity of the electroweak theory%
\refmark\ghw.
Moreover, if we focus on a given CP-even Higgs state, we note that
its couplings to $VV$ and $\ha V$ cannot be simultaneously
suppressed, since eqs.~\vvcoup--\hvcoup\ imply that
$$
  g^2_{h ZZ} + 4m^2_Z g^2_{hA^0Z} = {g^2m^2_Z\over
        \cos^2\theta_W}\,,\eqn\hxi
$$
for $h=\hl$ or $\hh$.  Similar considerations also hold for
the coupling of $\hl$ and $\hh$ to $W^\pm H^\mp$.  We can summarize
the above results by noting that the coupling of $\hl$ and $\hh$ to
vector boson pairs or vector--scalar boson final states is proportional
to either $\sin(\beta-\alpha)$ or $\cos(\beta-\alpha)$ as indicated
below.
$$\vbox{\settabs 2 \columns
\+\us{$\cos(\beta-\alpha)$}&  \us{$\sin(\beta-\alpha)$}\cr
\vskip2pt
\+    $H^0W^+W^-$&    $h^0W^+W^-$ \cr
 \+   $H^0ZZ$&        $h^0ZZ$ \cr
  \+  $ZA^0h^0$&      $ZA^0H^0$ \cr
   \+ $W^\pm H^\mp h^0$& $W^\pm H^\mp H^0$\cr }
\eqn\littletable$$

The 3-point and 4-point Higgs self-couplings depend on the
parameters of the
two-Higgs-doublet potential [eq.~\pot].  The Feynman rules for
the trilinear Higgs vertices are listed below; the
rule for the $ABC$ vertex is denoted by $ig\ls{ABC}$.
For completeness, $R$-gauge Feynman rules
involving the Goldstone bosons ($G^\pm$ and $G^0$)
are also listed.

{
$$\eqalign{%
g\ls{h^0A^0A^0} &=
   {2\mw\over g}\bigl[ \lambda_1\sb^2\cb\sa
   -\lambda_2\cb^2\sb\ca -\widetilde\lambda_3
   (\sb^3\ca-\cb^3\sa) +2\lambda_5\sba\cr
&\qquad -\lambda_6\sb\big(\cb\sab+\sa
c_{2\beta}\big)-\lambda_7\cb\big(\ca c_{2\beta}-\sb\sab\big)\bigr]\,,\crr\crr
g\ls{H^0A^0A^0} &= {-2\mw\over g}\bigl[
   \lambda_1\sb^2\cb\ca+\lambda_2\cb^2\sb\sa+\widetilde\lambda_3
   (\sb^3\sa+\cb^3\ca) -2\lambda_5\cba\cr
&\qquad -\lambda_6\sb\big(\cb\cab+\ca
c_{2\beta}\big)+\lambda_7\cb\big(\sb\cab+\sa c_{2\beta}\big)\bigr]\,,\crr\crr
g\ls{h^0H^0H^0} &= {6\mw\over g}\bigl[
   \lambda_1\ca^2\cb\sa-\lambda_2\sa^2\sb\ca+\widetilde\lambda_3
   (\sa^3\cb-\ca^3\sb+\twothirds\sba)\cr
&\quad -\lambda_6\ca\big(\cb c_{2\alpha}-\sa\sab\big)
-\lambda_7\ca\big(\sb c_{2\alpha}+\ca\sab\big)\bigr]\,,\crr\crr
g\ls{H^0h^0h^0} &= {-6\mw\over g}\bigl[
   \lambda_1\sa^2\cb\ca+\lambda_2\ca^2\sb\sa+\widetilde\lambda_3
   (\sa^3\sb+\ca^3\cb-\twothirds\cba)\cr
&\quad -\lambda_6\sa\big(\cb c_{2\alpha}+\ca\cab\big)
+\lambda_7\ca\big(\sb c_{2\alpha}+\sa\cab\big)\bigr]\,,\crr\crr
g\ls{h^0h^0h^0} &= {6\mw\over g}\bigl[
   \lambda_1\sa^3\cb-\lambda_2\ca^3\sb+\widetilde\lambda_3
   \sa\ca\cab\cr
&\quad -\lambda_6\sa^2\big(3\ca\cb-\sa\sb\big)
+\lambda_7\ca^2\big(3\sa\sb-\ca\cb\big)\bigr]\,,\crr\crr
g\ls{H^0H^0H^0} &= {-6\mw\over g}\bigl[
   \lambda_1\ca^3\cb+\lambda_2\sa^3\sb+\widetilde\lambda_3
   \sa\ca\sab\cr
&\quad +\lambda_6\ca^2\big(3\sa\cb+\ca\sb\big)
+\lambda_7\sa^2\big(3\ca\sb+\sa\cb\big)\bigr]\,,\crr\crr
g\ls{h^0H^+H^-} &=g\ls{h^0A^0A^0}-{2\mw\over g}
   \big(\lambda_5-\lambda_4\big)\sba\,,\crrr
g\ls{H^0H^+H^-} &=g\ls{H^0A^0A^0}-{2\mw\over g}
   \big(\lambda_5-\lambda_4\big)\cba\,,\cr}
\eqn\defghaa$$    \goodbreak
\vskip15pt\noindent
where I have used the notation
$$
\widetilde\lambda_3\equiv\lambda_3+\lambda_4+\lambda_5\,.
\eqn\lthree$$
It is interesting to note that couplings of the charged Higgs bosons
satisfy relations analogous to that of $\mhpm$ given in eq.~\mamthree.

Feynman rules for three-point Higgs vertices that involve Goldstone
bosons assume much simpler forms
$$\eqalign{
g\ls{h^0G^0G^0} &=
   {-g\over 2\mw} \mhl^2\sin(\beta-\alpha)\,,\crrr
g\ls{H^0G^0G^0} &=
   {-g\over 2\mw} \mhh^2\cos(\beta-\alpha)\,,\crrr
g\ls{h^0G^+G^-} &=g\ls{h^0G^0G^0}\,,\crrr
g\ls{H^0G^+G^-} &=g\ls{H^0G^0G^0}\,,\crrr
g\ls{h^0A^0G^0} &= {-g\over 2\mw}(\mhl^2-\mha^2)\cos(\beta-\alpha)\,,\crrr
g\ls{H^0A^0G^0} &= {g\over 2\mw}(\mhh^2-\mha^2)\sin(\beta-\alpha)\,,\crrr
g\ls{h^0H^\pm G^\mp} &=
{g\over 2\mw}(\mhpm^2-\mhl^2)\cos(\beta-\alpha)\,,\crrr
g\ls{H^0H^\pm G^\mp} &=
{-g\over 2\mw}(\mhpm^2-\mhh^2)\sin(\beta-\alpha)\,,\crrr
g\ls{A^0H^\pm G^\mp} &=
{\pm g\over 2\mw}(\mhpm^2-\mha^2)\,.\cr}\eqn\goldstonerules$$
} \vskip6pt\noindent
In the rule for the $\ha H^\pm G^\mp$ vertex, the sign corresponds to
$H^\pm$ entering the vertex and $G^\pm$ leaving the vertex.
One can easily check that if tree-level MSSM relations are imposed
on the $\lambda_i$, Higgs masses, and angles $\alpha$ and $\beta$,
one recovers the MSSM Feynman rules listed in Appendix A of ref.~[\hhg].
The Feynman rules for the 4-point Higgs vertices are rather
tedious in the general two-Higgs-doublet
model and will not be given here.

\REF\gw{S. Glashow and S. Weinberg, {\sl Phys. Rev.} {\bf D15}
(1977) 1958.}
The Higgs couplings
to fermions are model dependent, although their form is often
constrained by discrete symmetries that are imposed in order to
avoid tree-level flavor changing neutral currents mediated by Higgs
exchange\refmark\gw.
An example of a model that respects this constraint is one in
which one Higgs doublet (before symmetry
breaking) couples exclusively to down-type fermions and the other
Higgs doublet couples exclusively to up-type
fermions.  This is
the pattern of couplings found in the minimal supersymmetric model
(MSSM).  The results
in this case are as follows.  The couplings
of the neutral Higgs bosons to $f\bar f$
relative to the Standard Model
value, $gm_f/2\mw$, are given by (using 3rd family notation)
$${
\eqalign{
\hh t \bar t:&~~~{\sina\over\sinb}\crrr
\hl t \bar t:&~~~{\cosa\over\sinb}\crrr
\ha t \bar t:&~~~\gamma_5{\cot\beta}\cr }
\qquad\qquad
\eqalign{
\hh b \bar b:&~~~{\cosa\over\cosb}\crrr
\hl b \bar b:&~~~{-\sina\over\cosb}\crrr
\ha b \bar b:&~~~\gamma_5{\tanb}, \cr }
}\eqn\qqcouplings$$
\vskip5pt\noindent
(the $\gamma_5$ indicates a pseudoscalar coupling), and the
charged Higgs boson coupling is given by
$$g_{H^- t\bar b}={g\over{2\sqrt{2}\mw}}\
[m_t\cot\beta (1+\gamma_5)+m_b\tan\beta (1-\gamma_5)].
\eqn\hpmqq$$

\vskip1in
\APPENDIX{B}{B:~~Renormalization Group Equations}
\medskip
\REF\rgecollection{T.P. Cheng, E. Eichten and L.-F. Li, {\sl Phys. Rev.}
{\bf D9} (1974) 2259; K. Inoue, A. Kakuto, and Y. Nakano,
{\sl Prog. Theor. Phys.} {\bf 63} (1980) 234; H. Komatsu,
{\sl Prog. Theor. Phys.} {\bf 67} (1982) 1177; K. Inoue, A. Kakuto,
H. Komatsu and S. Takeshita,
{\sl Prog. Theor. Phys.} {\bf 67} (1982) 1889; {\bf 68} (1982) 927
[E: {\bf70} (1983) 330] {\bf 71} (1984) 413; C. Hill, C.N. Leung and
S. Rao, {\sl Nucl. Phys.} {\bf B262} (1985) 517.}
In this Appendix, I have collected
the one-loop renormalization group equations (RGEs)
that are needed in the analysis presented in this paper%
\refmark{\rgecollection,\bagger,\llog}.
Schematically, the RGEs at one-loop take the form
$$
{dp_i\over dt}=\beta_i(p_1,p_2,..)\,,\qquad
\hbox{ where}~t\equiv\ln\,\mu^2\,,\eqn\rges$$
where $\mu$ is the energy scale, and
the parameters $p_i$ stand for
the Higgs boson self-couplings $\lambda_i$ ($i=1\ldots 7$),
the squared Yukawa couplings $h_f^2$
($f=t$, $b$ and $\tau$; the two lighter generations can be neglected),
and the squared gauge couplings $g_j^2$ ($j=$3, 2, 1) corresponding to
SU(3)$\times$SU(2)$\times$U(1) respectively.  The $g_j$ are
normalized such that they are equal at the grand unification
scale.  It is also convenient to define
$$
g\equiv g\ls2\,,\qquad\qquad g\pri\equiv \sqrt{\threefifths} g\ls1\,,
\eqn\gaugecoups$$
where $g$ and $g\pri$ are normalized in the usual way for low-energy
electroweak physics, {\it i.e.} $\tan\theta_W=g\pri/g$.

I now list the $\beta$-functions required for the analysis presented
in this paper.  Two cases will be given, depending on whether $\mu$ is
above or below the scale of supersymmetry breaking, $\msusy$.
\medskip   \goodbreak
1. $\mu>\msusy$
$$\eqalign{%
\beta_{h_t^2}&={h_t^2\over16\pi^2}\left[6
  h_t^2+ h_b^2-\sixteenthirds g_3^2-3g^2-\thirteenninths g'^2\right]\crr
\beta_{h_b^2}&={h_b^2\over16\pi^2}\left[6 h_b^2+h_t^2
   +h_{\tau}^2-\sixteenthirds g_3^2-3g^2-\sevenninths g'^2\right]\crr
\beta_{h_{\tau}^2}&={h_{\tau}^2\over16\pi^2}\left[4
   h_{\tau}^2+3 h_b^2-3g^2-3g'^2\right]\crr
\beta_{g'^2}&={g'^4\over48\pi^2}\Big[10N_g+\threehalf N_H\Big]\crr
\beta_{g^2}&={g^4\over48\pi^2}\Big[6N_g+\threehalf N_H-18\Big]\crr
\beta_{g_3^2}&={g_3^4\over48\pi^2}\Big[6N_g-27\Big]\,.\cr}
\eqn\betagg$$
Here $N_g=3$ is the number of generations,
$N_H=2$ is the number of scalar doublets,
and the Higgs-fermion Yukawa couplings are given by
$$\eqalign{%
h_t&={gm_t\over \sqrt2 m_W \sinb}\,,\crr
h_{d_i}&={gm_d\over \sqrt2 m_W \cosb}\,,\qquad (d_i =
b,\tau)\,.\cr}\eqn\yukawas$$
\par        \nobreak
2. $\mu<\msusy$
$$\eqalign{%
\beta_{h_t^2}      &={h_t^2\over16\pi^2}\big[\ninehalf
h_t^2+\half h_b^2-8g_3^2-\ninefourth
g^2-\seventeentwelfth g'^2\big]\crr
\beta_{h_b^2}      &={h_b^2\over 16\pi^2}\big[\ninehalf h_b^2+\half
h_t^2+h_{\tau}^2-8g_3^2-\ninefourth g^2-\fivetwelfth
g'^2\big]\crr
\beta_{h_{\tau}^2} &={h_{\tau}^2\over16\pi^2}\big[\fivehalf
h_{\tau}^2+3h_b^2-\ninefourth g^2-\fifteenfourth
g'^2\big]\crr
%\eqn\rgeh$$
%$$\eqalign{
\beta_{g'^2}  &={g'^4\over48\pi^2}\Big[\twentythirds N_g+\half
N_H\Big]\crr
\beta_{g^2}   &={g^4\over48\pi^2}\Big[4N_g+\half N_H-22\Big]\crr
\beta_{g_3^2} &={g_3^4\over48\pi^2}\Big[4N_g-33\Big]\,.\cr}
\eqn\betagg$$
In writing down the
RGEs for the Higgs-fermion Yukawa couplings, I have assumed that
the Higgs-fermion interaction is the same as in the MSSM; namely,
$\Phi_1$ [$\Phi_2$] couples exclusively to down-type
[up-type] fermions.

Finally, the RGEs for the
Higgs self-couplings
%of the general two-Higgs doublet model (with
%the Higgs-fermion couplings as specified above).
%First, I need to define the anomalous dimensions of the two Higgs
%fields:
%The $\beta$-functions for the Higgs self-couplings
in the general CP-conserving
non-supersymmetric two-Higgs-doublet model (with
the Higgs-fermion couplings as specified in Appendix A)
are  given by
$$\eqalign{%
\beta_{\lambda_{1}} &={1\over16\pi^2}
   \bigg\{ 6\lambda^2_{1}+2\lambda_3^2+2\lambda_3\lambda_4+
   \lambda_4^2+\lambda_5^2+12\lambda_{6}^2 \bigg.\cr
&\hskip2cm \bigg.        +\threeighth
       \big[2g^4+(g^2+g'^2)^2\big]-2\sum_i N_{ci}h_{d_i}^4\bigg\}
      -2\lambda_{1}\gamma_{1}\crr
\beta_{\lambda_{2}} &={1\over16\pi^2}
   \bigg\{ 6\lambda^2_{2}+2\lambda_3^2+2\lambda_3\lambda_4+
   \lambda_4^2+\lambda_5^2+12\lambda_{7}^2 \bigg.\cr
&\hskip2cm \bigg. +\threeighth
      \big[2g^4+(g^2+g'^2)^2\big]-2\sum_i N_{ci}h_{u_i}^4\bigg\}
      -2\lambda_{2}\gamma_{2}\crr
\beta_{\lambda_3} &={1\over16\pi^2}
   \bigg\{ (\lambda_1+\lambda_2)(3\lambda_3+\lambda_4)+2\lambda_3^2
   +\lambda_4^2+\lambda_5^2+2\lambda_6^2+2\lambda_7^2
   +8\lambda_6\lambda_7 \bigg.\cr
&\hskip2cm +\threeighth   \bigg.
      \big[2g^4+(g^2-g'^2)^2\big]-2\sum_i N_{ci}h_{u_i}^2h_{d_i}^2\bigg\}
      -\lambda_3(\gamma_1+\gamma_2)\crr
\beta_{\lambda_4}  &={1\over16\pi^2}
   \Big[ \lambda_4(\lambda_1+\lambda_2+4\lambda_3+2\lambda_4)+
   4\lambda_5^2+5\lambda_6^2+5\lambda_7^2+2\lambda_6\lambda_7\Big.\crr
&\hskip2cm +\threehalf       \Big.
      g^2g'^2+2\sum_i N_{ci}h_{u_i}^2h_{d_i}^2\Big]
      -\lambda_4(\gamma_1+\gamma_2)\crr
\beta_{\lambda_5} &={1\over16\pi^2}
   \Big[\lambda_5(\lambda_1+\lambda_2+4\lambda_3+6\lambda_4)
   +5\big(\lambda_6^2+\lambda_7^2\big)+2\lambda_6\lambda_7\Big]
   -\lambda_5(\gamma_1+\gamma_2)\crr
\beta_{\lambda_6} &={1\over16\pi^2}
   \Big[\lambda_6(6\lambda_1+3\lambda_3+4\lambda_4+5\lambda_5\big)
   +\lambda_7\big(3\lambda_3+2\lambda_4+\lambda_5\big)\Big]
   -\half\lambda_6(3\gamma_1+\gamma_2)\crr
\beta_{\lambda_7} &={1\over16\pi^2}
   \Big[\lambda_7(6\lambda_2+3\lambda_3+4\lambda_4+5\lambda_5\big)
   +\lambda_6\big(3\lambda_3+2\lambda_4+\lambda_5\big)\Big]
   -\half\lambda_7(\gamma_1+3\gamma_2)\,. \crr}
\eqn\betal$$
In eq.~\betal, the anomalous dimensions of the two
Higgs fields are given by
$$\eqalign{%
\gamma_{1} &={1\over 64\pi^2}
   \Big[9g^2+3g'^2-4\sum_i N_{ci}h_{d_i}^2\Big]\,,\crr
\gamma_{2} &={1\over 64\pi^2}
    \Big[9g^2+3g'^2-4\sum_i N_{ci}h_{u_i}^2\Big]\,,\cr}
\eqn\wavez$$
where the sum over $i$ is taken over three generations of quarks
(with $N_c=3$) and leptons (with $N_c=1$).
\endpage

\APPENDIX{C}{C:~~MSSM Higgs Sector Contributions}%\break
\vskip-4pt
\centerline{\fourteenpoint to the $S,\ T$ and $U$ Parameters}
\medskip
\REF\passarino{G. Passarino and M. Veltman, {\sl Nucl. Phys.}
{\bf B160} (1979) 151.}
\REF\bjdrell{J.D. Bjorken and S.D. Drell, {\it Relativistic Quantum
Mechanics} (McGraw-Hill, New York, 1964).}

In this Appendix, I record the exact one-loop contributions of the
MSSM Higgs bosons to $\delta S,\ \delta T$ and $\delta U$
[see  eq.~\stsusy].
These contributions are defined relative to the Standard Model in
which the Standard Model Higgs boson mass is taken equal to
$\mhl$\refmark\hehth.
First, for $\delta S$ and $\delta U$ I find
$$\eqalign{
\delta S&=  {1\over\pi\mz^2}\Biggl\{
\sin^2(\beta-\alpha)
{\cal B}_{22}(\mz^2;\mhh^2,\mha^2)-{\cal B}_{22}(\mz^2;\mhpm^2,
\mhpm^2)\cr
& +\cos^2(\beta-\alpha)\biggl[{\cal B}_{22}
(\mz^2;\mhl^2,\mha^2)+{\cal B}_{22}(\mz^2;\mz^2,\mhh^2)
-{\cal B}_{22} (\mz^2;\mz^2,\mhl^2)\cr
&\qquad\qquad -\mz^2{\cal B}_0(\mz^2;\mz^2,\mhh^2)
+\mz^2{\cal B}_0(\mz^2;\mz^2,\mhl^2)\biggr]\Biggr\}\,.\crr
\delta U&= -\delta S+{1\over\pi\mz^2}\Biggl\{
{\cal B}_{22}(\mw^2;\mha^2,\mhpm^2)-2{\cal B}_{22}(\mw^2;\mhpm^2,
\mhpm^2)\crr
& +\sin^2(\beta-\alpha)
{\cal B}_{22}(\mw^2;\mhh^2,\mhpm^2)  \crr
& +\cos^2(\beta\!-\!\alpha)\biggl[{\cal B}_{22}
(\mw^2;\mhl^2,\mhpm^2)\!+\!{\cal B}_{22}(\mw^2;\mw^2,\mhh^2)
\!-\!{\cal B}_{22} (\mw^2;\mw^2,\mhl^2)\cr
&\qquad\qquad -\mw^2{\cal B}_0(\mw^2;\mw^2,\mhh^2)
+\mw^2{\cal B}_0(\mw^2;\mw^2,\mhl^2)\biggr]\Biggr\}\,.\cr}
\eqn\esshiggs$$
The following notation has been introduced for the various
loop integrals
$${\cal B}_{22}(q^2;m_1^2,m_2^2)\equiv B_{22}(q^2;m_1^2,m_2^2)-
B_{22}(0;m_1^2,m_2^2)\,,\eqn\calbtwo$$
$${\cal B}_{0}(q^2;m_1^2,m_2^2)\equiv B_{0}(q^2;m_1^2,m_2^2)-
B_{0}(0;m_1^2,m_2^2)\,,\eqn\calbzero$$
and $B_{22}$ and $B_0$ are defined according to ref.~[\passarino]
(up to an overall sign in some cases since I use the Bjorken and Drell
metric\refmark\bjdrell).  Explicitly,
$$B_{22}(q^2;m_1^2,m_2^2)=\fourth(\Delta+1)\left[m_1^2+m_2^2
-\third q^2\right]-\half\int_0^1\,dx\,X\ln(X-i\epsilon)
\,,\eqn\beetwo$$
\vskip-5pt
$$B_0(q^2;m_1^2,m_2^2)=\Delta-\int_0^1\,dx\,\ln(X-i\epsilon)
\,,\eqn\beezero$$
where
$$X\equiv m_1^2x+m_2^2(1-x)-q^2 x(1-x)\eqn\xdefinition$$
and $\Delta$ is the regulator of dimensional regularization defined by
$$\Delta={2\over 4-n}+\ln(4\pi)+\gamma\,,\eqn\deltadef$$
$n$ is the number of space-time
dimensions and $\gamma$ is Euler's constant.  Of course, in the
calculation of physical observables, terms proportional to $\Delta$ must
exactly cancel.  The following two
relations are particularly useful:
$$4B_{22}(0;m_1^2,m_2^2)=F(m_1^2,m_2^2)+A_0(m_1^2)+A_0(m_2^2)\,,
\eqn\bzerotwo$$
$$B_0(0;m_1^2,m_2^2)={A_0(m_1^2)-A_0(m_2^2)\over m_1^2-m_2^2}\,,
\eqn\bzerozero$$
where
$$F(m_1^2,m_2^2)\equiv\half(m_1^2+m_2^2)-{m_1^2 m_2^2\over m_1^2-
m_2^2}\,\ln\left({m_1^2\over m_2^2}\right)\,,\eqn\effdef$$
$$A_0(m^2)\equiv m^2(\Delta+1-\ln m^2)\,.\eqn\azero$$
Finally, consider $\delta\rho=\alpha \delta T$.  I find:
$$\eqalign{\delta T=&{1\over 16\pi\mw^2 s\ls{W}^2}\Biggl\{
F(\mhpm^2,\mha^2)+\sin^2(\beta-\alpha)\left[F(\mhpm^2,\mhh^2)-
F(\mha^2,\mhh^2)\right]\crr &\qquad\qquad
+\cos^2(\beta-\alpha)\left[F(\mhpm^2,\mhl^2)
-F(\mha^2,\mhl^2)+F(\mw^2,\mhh^2)\right.\crr &\left.
\qquad\qquad\qquad-F(\mw^2,\mhl^2)
-F(\mz^2,\mhh^2)+F(\mz^2,\mhl^2)\right]\crr & \qquad\qquad
+4\mz^2\left[B_0(0;\mz^2,\mhh^2)-B_0(0;\mz^2,\mhl^2)\right]\crr &
\qquad\qquad
-4\mw^2\left[B_0(0;\mw^2,\mhh^2)-B_0(0;\mw^2,\mhl^2)\right]\Biggr\}
\,,\cr}\eqn\teehiggs$$
where $s\ls{W}\equiv\sin\theta_W$.

Note that these expressions for $\delta S,\ \delta T$ and $\delta U$
are valid for an
arbitrary two-Higgs doublet extension of the Standard
Model\refmark{\higgsrad--\higgsradtwo}.
In the MSSM, tree-level relations exist among the Higgs masses and
angles $\alpha$ and $\beta$ as discussed in section 3.  By virtue of
these relations, the numerical values of $\delta S$ and
$\delta T$ in the MSSM are much smaller than 1.  For example, by
taking $\mha\gg\mz$ and applying the various MSSM tree-level
relations, one easily obtains the asymptotic results quoted in
eqs.~\shiggsapprox\ and \thiggsapprox.  Moreover, one can check
that $\delta U \ll \delta S,\ \delta T$.

\refout
\endpage
\figout
\bye